%% file: main.tex
\documentclass[sigconf, screen, nonacm=true]{acmart}
\AtBeginDocument{%
  \providecommand\BibTeX{{%
    \normalfont B\kern-0.5em{\scshape i\kern-0.25em b}\kern-0.8em\TeX}}}

\copyrightyear{2023}
\acmYear{2023}
\setcopyright{acmlicensed}
\acmConference[ISCA '23] {Proceedings of the 50th Annual International Symposium on Computer Architecture}{June 17--21, 2023}{Orlando, FL, USA.}
\acmBooktitle{Proceedings of the 50th Annual International Symposium on Computer Architecture (ISCA '23), June 17--21, 2023, Orlando, FL, USA}
\acmPrice{15.00}
\acmISBN{979-8-4007-0095-8/23/06}
\acmDOI{10.1145/3579371.3589042}
\usepackage{graphicx}
\usepackage{textcomp}

\newcommand{\subheading}[1]{\vspace{0.03in}{\noindent \textbf{#1.}}}

\usepackage{tikz}
\usepackage[most]{tcolorbox}
\usepackage{hyperref}
\usepackage[linesnumbered,ruled]{algorithm2e}
\usepackage{multirow}
\usepackage{subfigure}
\usepackage{colortbl}
\newcommand*\circled[1]{\tikz[baseline=(char.base)]{
        \node[shape=circle,fill=black,draw,inner sep=0.15pt] (char) {\color{white}\fontfamily{phv}\selectfont{#1}};}}
\newcommand{\ours}{{\sc herqules}}
\newcommand{\ourspace}{{\sc herqules }}
\newcommand{\rev}[1]{\textcolor{black}{#1}}
\begin{document}

\title{Scaling Qubit Readout with Hardware Efficient Machine Learning Architectures}

\author{Satvik Maurya}
\email{smaurya@wisc.edu}
\orcid{}
\affiliation{%
  \institution{University of Wisconsin-Madison}
  \streetaddress{}
  \city{Madison}
  \state{WI}
  \country{USA}
  \postcode{}
}

\author{Chaithanya Naik Mude}
\email{cmude@wisc.edu}
\orcid{}
\affiliation{%
  \institution{University of Wisconsin-Madison}
  \streetaddress{}
  \city{Madison}
  \state{WI}
  \country{USA}
  \postcode{}
}

\author{William D. Oliver}
\email{william.oliver@mit.edu}
\orcid{}
\affiliation{%
  \institution{Massachusetts Institute of Technology}
  \streetaddress{}
  \city{Cambridge}
  \state{MA}
  \country{USA}
  \postcode{}
}

\author{Benjamin Lienhard}
\email{blienhard@princeton.edu}
\orcid{}
\affiliation{%
  \institution{Princeton University}
  \streetaddress{}
  \city{Princeton}
  \state{NJ}
  \country{USA}
  \postcode{}
}

\author{Swamit Tannu}
\email{swamit@cs.wisc.edu}
\orcid{}
\affiliation{%
  \institution{University of Wisconsin-Madison}
  \streetaddress{}
  \city{Madison}
  \state{WI}
  \country{USA}
  \postcode{}
}


\begin{abstract}
Reading a qubit is a fundamental operation in quantum computing. It translates quantum information into classical information enabling subsequent classification to assign the qubit states `0' or `1'.
Unfortunately, qubit readout is one of the most error-prone and slowest operations on a superconducting quantum processor. On state-of-the-art superconducting quantum processors, readout errors can range from 1-10\%.
These errors occur for various reasons -- crosstalk, spontaneous state transitions, and excitation caused by the readout pulse. The error-prone nature of readout has resulted in significant research to design better discriminators to achieve higher qubit-readout accuracies. High readout accuracy is essential for enabling high fidelity for near-term noisy quantum computers and error-corrected quantum computers of the future.

Prior works have used machine-learning-assisted single-shot qubit-state classification, where a deep neural network was used for more robust discrimination by compensating for crosstalk errors. 
However, the neural network size can limit the scalability of systems, especially if fast hardware discrimination is required. This state-of-the-art baseline design cannot be implemented on off-the-shelf FPGAs used for the control and readout of superconducting qubits in most systems, which increases the overall readout latency as discrimination has to be performed in software.

In this work, we propose \ours, a scalable approach to improve qubit-state discrimination by using a hierarchy of matched filters in conjunction with a significantly smaller and scalable neural network for qubit-state discrimination. We achieve substantially higher readout accuracies (16.4\% relative improvement) than the baseline with a scalable design that can be readily implemented on off-the-shelf FPGAs. We also show that \ourspace is more versatile and can support shorter readout durations than the baseline design without additional training overheads.  
\end{abstract}






\maketitle

\input{inputs/1intro} 
\input{inputs/2background} 

\input{inputs/3problem}
\input{inputs/4design-1} 
\input{inputs/5design-2}
\input{inputs/6Methodology} 
\input{inputs/7Evaluation} 
\input{inputs/8Discussion}
\input{inputs/9Conclusion} 

\section*{Acknowledgements}
We thank the anonymous reviewers for their feedback. This research was supported by the National Science Foundation award: 2212232 and the Vice Chancellor Office for Research and Graduate Education at the University of Wisconsin–Madison with funding from the Wisconsin Alumni Research Foundation.

\rev{
\vspace{-0.2in}
\section*{Appendix A}
\label{sec:appA}
\subheading{Matched filters} 
Qubit readout accuracy is highly dependent on the Signal-to-Noise Ratio (SNR) of the acquired readout signal. Matched filters are optimal for readout because they capture the time-varying SNR of the readout signal by having individual weights for every time-step of the readout signal. We encourage readers to peruse prior works~\cite{Lienhard2022,ryan2015tomography} for more elaborate explanations on matched filters for qubit readout. The weights, or envelope, of a matched filter is derived by dividing the mean with the variance of the difference vector of the ground (`0') and excited (`1') state traces (Tr) (one envelope each for the I and Q components), as described by the equation below:
}
$${ MF~Envelope = \frac{mean(Tr_0 - Tr_1)}{var(Tr_0 - Tr_1)}}$$

\bibliographystyle{ACM-Reference-Format}
\balance
\bibliography{refs}

\end{document}

%% file: inputs/1intro.tex




\section{Introduction}


Quantum computers leverage fundamental quantum-mechanical properties of their constituent quantum bits (qubits) to gain a computational advantage for a specific class of complex problems. Today, scientists and engineers are racing to build larger, more-reliable quantum computers and demonstrate their effectiveness at evaluating increasingly complex quantum algorithms. Quantum hardware has two primary components: Qubits, which hold the quantum information, and a control computer that manipulates this information to orchestrate the execution of quantum programs. The control computer is further divided into a qubit readout and control pipeline. The control pipeline sends precise gate pulses to qubits, whereas the readout pipeline measures the qubits. Control and readout are performed on existing quantum computers with hundreds of qubits using FPGAs and signal generators.




Readout is a fundamental operation in quantum computing. It converts quantum information into classical information represented by the computational space (`0' and `1'). Readout of superconducting qubits involves three stages -- (1) query the qubit state using a readout pulse, (2) acquire the response from the qubit, and (3) discriminate the acquired signal to infer the qubit state. Unfortunately, qubit readout is among the most error-prone and slowest operations on superconducting quantum processors. On state-of-the-art cloud-based quantum processors, readout errors can exceed 10\% for some qubits~\cite{ibm_systems} with readout durations generally surpassing 300ns. In quantum processors used to demonstrate quantum supremacy~\cite{Arute2019}, readout errors can be as high as 9\%~\cite{weber} with durations exceeding 1$\mu$s.




Due to the complexity of the readout process, which involves multiple analog components like filters and amplifiers, reducing readout errors cannot be achieved by merely optimizing a single part in the pipeline. Improvements to readout hardware, such as Purcell filters and Josephson Traveling Wave Parametric Amplifiers~\cite{Macklin2015, Heinsoo2018, Sete2015} have improved qubit-readout accuracy. Such hardware enhancements are beyond the scope of this paper. In contrast, statistical methods that mitigate readout errors on a distribution of readout results~\cite{Bravyi2021} have been proposed to help applications of the Noisy Intermediate Scale Quantum (NISQ) era, where the same program is run thousands of times to generate a probability distribution to reveal the most likely result. However, such statistical methods do not improve the readout accuracy for single-shot readout, i.e., the accuracy of a single measurement. In this work, we focus on \emph{improving single-shot qubit-state discrimination}. 



Apart from improving the analog components of the readout hardware, single-shot qubit-readout accuracy can be improved by designing better discriminators. The state-of-the-art deep neural network-assisted discriminator demonstrated by Lienhard et al.~\cite{Lienhard2022} achieved a significantly higher readout accuracy for a group of frequency-multiplexed qubits compared to other discriminators, such as matched filters and support vector machines. This was achieved by using the digitized readout signal at the intermediate frequencies as an input to a large Feed Forward Neural Network (FNN) and training it to classify the state of all qubits. Using the readout signal directly from the analog-to-digital converter without any digital preprocessing, the FNN effectively utilizes all available information during each measurement to achieve better qubit-readout accuracy. However, such a discriminator architecture introduces a substantial overhead -- the size of the neural network prevents it from being implemented in a scalable manner on programmable hardware like FPGAs, which are used as control hardware for most quantum computers based on superconducting qubits today. Thus, the discriminator must be implemented in software, and the readout signal has to be transferred from the control hardware to software every time a qubit measurement is performed.


\begin{figure}
    \centering
    \includegraphics[width=\linewidth]{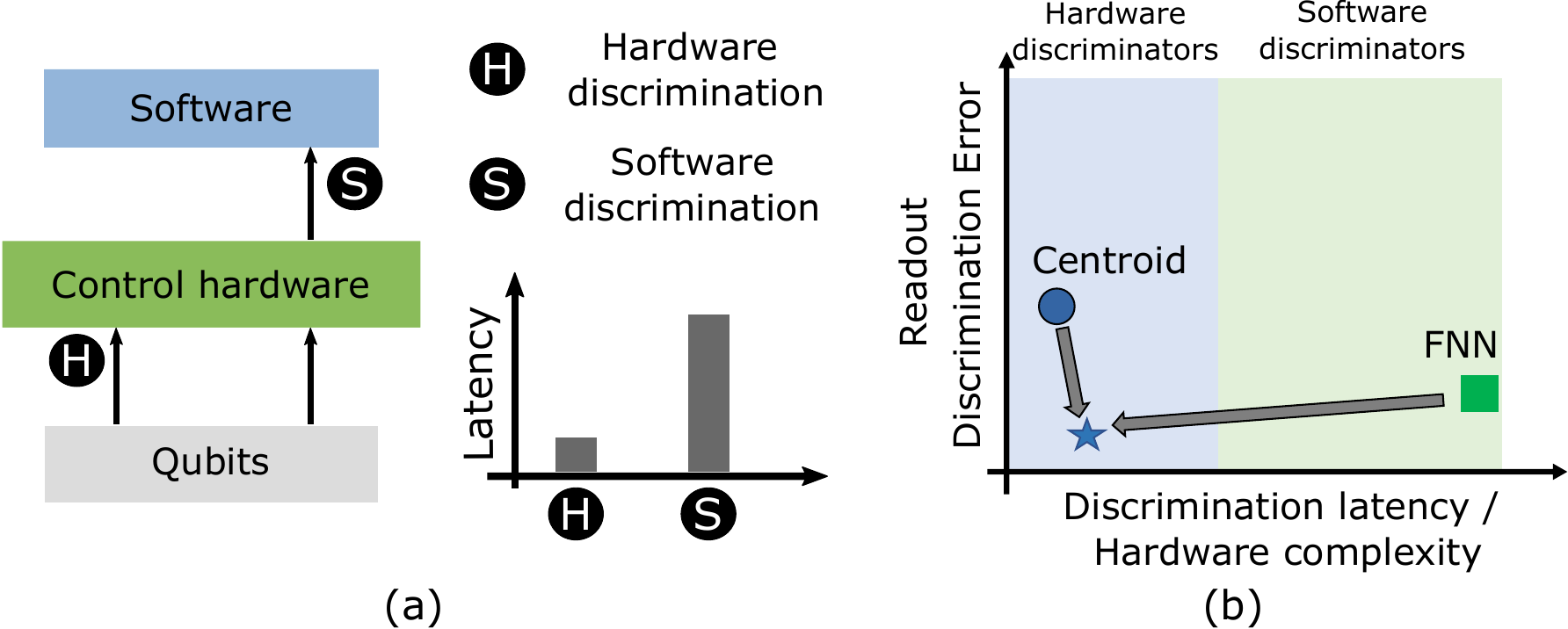}
    \vspace{-0.2in}
    \caption{(a) Latency trade-off between hardware and software discrimination; (b) \ours~goal: To reduce readout errors with low hardware complexity. 
    }
    \label{fig:fig1}
    \vspace{-0.25in}
\end{figure}

Most discriminators implemented on hardware are relatively simple and involve some form of filtering to reduce the noisiness or dimensionality of the readout data received from the qubits. Filters such as (mode) matched filters~\cite{ryan2015tomography, Walter2017, Arute2019} are commonly used to reduce the digitized readout signal received from the ADC into a single value that can be used for discrimination. 
On the other hand, the deep neural network discriminator~\cite{Lienhard2022} demonstrated that there is a plethora of information in the unfiltered readout signal that can be used for more accurate discrimination. However, as shown in Figure~\ref{fig:fig1}(a), using a software discriminator introduces a significant latency overhead to the readout operation due to the expense of transferring the high-dimensional readout traces to software.


To navigate the trade-off between the advantages of using the intermediate-frequency readout time traces for discrimination and hardware complexity, we propose \ours~-- High-fidElity haRdware efficient QuBit readoUt using Low-DimEnsional TraceS. As shown in Figure~\ref{fig:fig1}(b), \ours~is designed to provide improved readout accuracy over simple hardware discriminators such as the centroid classifier used in cloud-based superconducting quantum processors~\cite{ibm_discriminators} while having a much lower hardware complexity and discrimination latency than the FNN discriminator~\cite{Lienhard2022}. \ours~does this by first using matched filters (MF), one per qubit, to reduce the dimensionality of the qubit-readout time traces comprising of many time bins -- each time bin represents a dimension -- and using the low-dimensional outputs of the MFs as inputs to a lightweight neural network. However, this alone does not ensure higher readout accuracies than the baseline FNN discriminator~\cite{Lienhard2022} as substantial information contained in the high-dimensional traces is lost after dimensionality reduction. To address this, we introduce another MF for each qubit that can detect qubit relaxations (1$\rightarrow$0 transitions) that occur during readout (here referred to as relaxation matched filter (RMF)). MFs are trained using labeled data, which are not trivial to generate for qubit relaxations. We thus also propose an efficient, semi-supervised method of labeling data to identify qubit relaxations to create training data for the RMF. Using two different MFs for each qubit, we provide the neural network with additional features to improve the qubit state inference. Our evaluations show that such a design outperforms the baseline FNN design by more than 1.4\% -- which corresponds to a 16.4\% relative improvement -- in terms of cumulative accuracy on the same five-qubit dataset while having less than 8\% LUT utilization on an FPGA. 

\ours~is also more conducive for enabling faster readout. Readout is the slowest operation on superconducting quantum processors used today~\cite{ibm_systems, weber, Heinsoo2018}. Long readout durations increase the error rate due to qubit relaxations, making finding the shortest readout duration essential for high accuracies~\cite{gambetta2007protocols}. Furthermore, the readout duration is a crucial aspect of the cycle time in quantum error correction~\cite{versluis2017scalable}, which implies that reducing readout durations also benefits long-term applications. For NISQ applications, using mid-circuit measurements in applications like Quantum Phase Estimation~\cite{Crcoles2021} also presents opportunities for lower readout durations to have a positive impact. In the baseline design, the entire readout time trace is used to train an FNN; thus, the architecture depends on the readout duration (1$\mu$s for the baseline). Any reduction in the readout duration results in a need to retrain the FNN with the new readout duration, making it challenging to support shorter readout durations adaptively. On the other hand, we propose a method to reduce the readout duration for (1) all qubits and (2) for specific qubits without requiring additional training. \ours~can be trained using data corresponding to the entire readout duration (1$\mu$s), while shorter readout durations can be used during inference. This is possible because using MFs makes the neural network agnostic to the actual readout duration. 
We find that the readout duration for every qubit can be reduced by 25\% and still have a better overall performance than the baseline. 

The contributions of this paper are summarized below:
\begin{itemize}
    \item We propose \ours, a discriminator architecture that has high readout accuracy at minimal hardware cost by using matched filters to discriminate the readout signal and detect qubit-state-relaxations that occur during readout.
    \item We propose a simple, semi-supervised method to label training data for relaxation matched filters.
    \item We propose a method to reduce the readout duration for all or some qubits for application-specific requirements without incurring the cost of additional training.
\end{itemize}







%% file: inputs/2background.tex
\section{Background}
\label{sec:background}

In this section, we explain qubit readout for superconducting architectures and various sources of errors that impact readout fidelity.

\subsection{Readout of superconducting qubits}
\begin{figure}[t]
    \centering
    \includegraphics[width=0.8\linewidth]{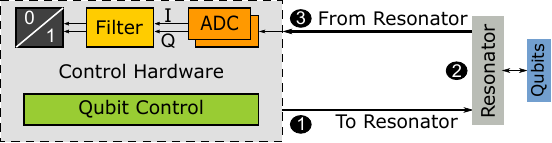}
    \vspace{-0.10in}
    \caption{The qubit readout pipeline comprises 1) readout-probe-tone generation, 2) acquisition of qubit-state-dependent phase shift, and 3) readout-signal processing.}
    \label{fig:readout_pipe}
    \vspace{-0.15in}
\end{figure}

Readout refers to the process of determining the state of a qubit. After a measurement, the qubit is typically in the ground state (`0') or excited state (`1'). Superconducting qubits are dispersively coupled to circuit elements called resonators which interact with the qubits for the purpose of readout. Readout involves multiple steps, as visualized in Figure~\ref{fig:readout_pipe}: \circled{1} a microwave pulse is driven into the resonator by the control hardware; \circled{2} the incident microwave pulse undergoes a phase shift that depends on the qubit state; \circled{3} the control hardware analyzes the phase shift to discriminate between the ground and excited states. Superconducting qubit readout involves multiple analog components along with robust digital signal processing to achieve appreciable accuracy. In this paper, we focus on improving step \circled{3} of the readout process, in terms of both accuracy and scalability.

\subsection{Qubit Readout Pipeline}

To determine whether a qubit is in state `0' or `1', the microwave signal transmitted or reflected off the readout resonator is typically frequency-downmodulated, digitized, and then processed in the control hardware. The incoming microwave signal is quadrature modulated. The In-phase(I) and Quadrature(Q) components of the signal are retrieved via analog mixing before they are digitized by two Analog-to-Digital Converters (ADCs). ADCs used for readout have sampling rates ranging from 250-1000 MSamples/sec, which makes the data generated by the ADCs for long readout pulses challenging to manage without additional dimensionality reduction. A (mode) matched filter~\cite{ryan2015tomography, Walter2017, Heinsoo2018, Arute2019} is used to reduce the incoming I and Q data stream from the ADCs to a single value. The filtered values are then used to discriminate the qubit state using techniques such as clustering or Support Vector Machines (SVMs)~\cite{Magesan}.

\begin{figure}[b]
    \centering
    \vspace{-0.23in}
    \includegraphics[width=\linewidth]{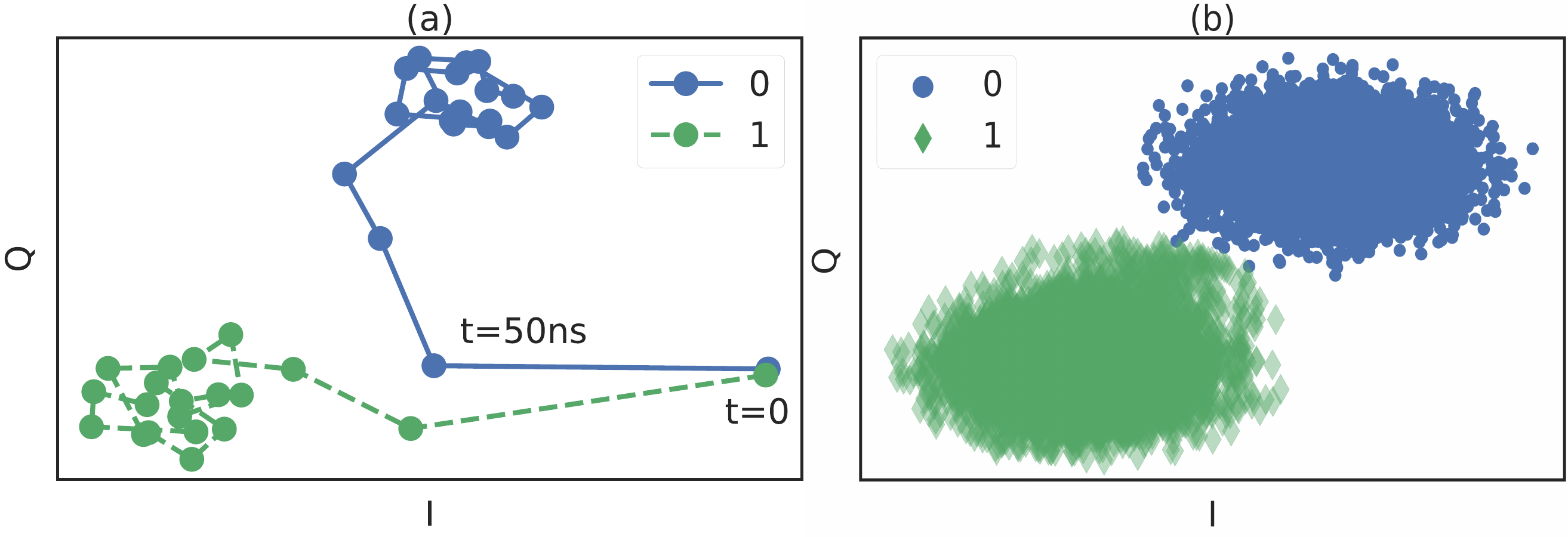}
    \vspace{-0.25in}
    \caption{(a) Evolution of readout traces for ground and excited states in the 2D I-Q plane -- the point annotated with `$t=0$' represents the first time bin of the readout trace, and every subsequent point comes later at a granularity of 50ns; (b) Mean values of multiple readout traces for both states showing the distinction between them.}
    \label{fig:trace_scatter}
\end{figure}
Figure~\ref{fig:trace_scatter} illustrates the signal generated during the readout process. In Figure~\ref{fig:trace_scatter}(a), the output of the ADCs (I and Q channels) is represented on a 2D plane showing how the qubit-readout trace evolves \rev{over time (each IQ-pair corresponds to a period of 50ns).} The temporal mean of one such trace corresponds to a single point in Figure~\ref{fig:trace_scatter}(b), which represents the Mean Trace Value ({\em MTV}). The MTV of a trace $Tr$ is given by the equation: $MTV=\frac{1}{len(Tr)}\sum_{t=0}^{len(Tr)}Tr(t)$. In the case of frequency-multiplexed readout where multiple qubits are measured using the same physical channel~\cite{Heinsoo2018, Walter2017, Chen2012, Arute2019, Lienhard2022}, the readout trace obtained from the ADCs can be demodulated to retrieve the traces of the constituent qubits. \rev{Digital demodulation of the readout signal requires the extraction of the readout signals for each qubit. For this work, digital demodulation is achieved by multiplying the frequency-multiplexed readout signal with an oscillating signal at a frequency specific to the readout resonator. The result is then averaged over intervals of 50ns.}

\begin{figure*}[t]
    \centering
    \includegraphics[width=0.9\textwidth]{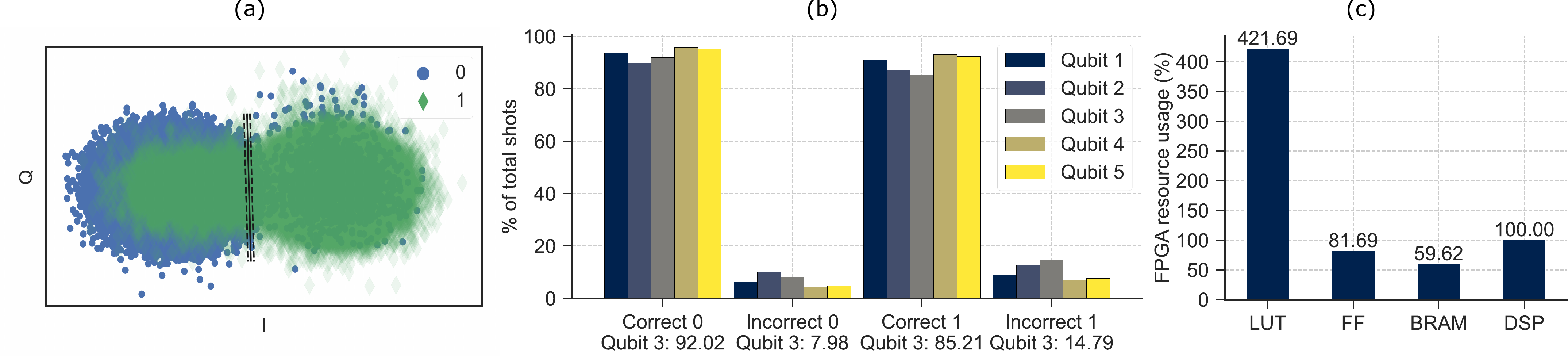}
    \vspace{-0.1in}
    \caption{(a) Scatter plot of 131,000 shots equally distributed between the 0 and 1 states for the fifth qubit of IBM Manila. An SVM discriminator trained using this data has a decision boundary that is heavily biased due to relaxation errors; (b) Single-shot measurements that can be discriminated correctly and incorrectly for both states of all qubits in IBM Manila; (c) Estimated hardware usage of a neural network 40\% of the size used in~\cite{Lienhard2022} for discriminating five qubits.}
    \label{fig:problem}
    \vspace{-0.15in}
\end{figure*}

\subsection{Readout errors}
High-fidelity readout is essential to enabling accurate computation on quantum computers. Unfortunately, qubit readout is among the most error-prone operations in state-of-the-art superconducting quantum processors. For example, on the 127-qubit IBM Washington quantum computer, readout error rates range from $\sim$0.1\% for some qubits to $>$10\% for other qubits~\cite{ibm_systems}. The primary sources of qubit-readout errors are described below.

\subheading{Relaxation errors} 
Most superconducting qubits cannot sustain the excited state (`1') for more than a few tens to hundreds of microseconds. Therefore, relaxations from the excited to the ground state are likely to occur during a long latency operation like readout~\cite{picot2008role}. 

\subheading{Crosstalk errors}
In the case of frequency-multiplexed readout, interactions (crosstalk) between the readout pulses can result in errors during classification. Frequency-multiplexed readout dramatically improves the scalability of systems since less hardware is needed for readout, which increases the importance of mitigating crosstalk errors.

\subheading{Excitation errors}
Errors due to undesired qubit excitations are similar to relaxation errors, except that they occur when the readout pulse excites a qubit~\cite{Walter2017} resulting in an incorrect measurement. 


\subheading{Errors due to enviromental noise}
Multiple filters and amplifiers are used to increase the signal-to-noise ratio (SNR) of the readout signal~\cite{Macklin2015, Sete2015, Neill2018}. Unfortunately, these components do not perfectly compensate for noise on the readout signal.

In the following sections, we describe how these errors affect readout fidelity and how they can be mitigated to make readout more robust in a scalable manner.


%% file: inputs/3problem.tex
\section{Scaling High-Fidelity Readout}
\label{sec:problem}
In this section, we describe the challenges in achieving scalable, high-fidelity single-shot qubit readout.


\subsection{Single-shot qubit readout fidelity}
Efforts to improve qubit-readout fidelities can be divided into two broad categories: (a) statistical error mitigation over an ensemble of measurement results, and (b) improving the readout accuracy per measurement (single-shot). Statistical error mitigation uses a trained error profile to reduce errors in the probability distribution of the measurement (obtained after running the same experiment for many shots). Measurement error mitigation~\cite{Bravyi2021, ibm_rem} is an example of such an approach, along with an alternative where neural networks are used to learn the error profile~\cite{Kim2022}. 

Methods to improve single-shot readout accuracy are beneficial for both near-term and long-term applications. Active error mitigation methods that use ancillary qubits for mitigating error via redundancy~\cite{Alexandrou2021, Funcke2022, Hicks2022} and methods that optimize the readout pipeline, such as deep learning assisted readout~\cite{Lienhard2022} are ways to improve the single-shot qubit readout accuracy. 
Since any improvements to the readout pipeline that result in better readout fidelity automatically benefit active readout error mitigation, we limit our discussion to techniques such as the Feed Forward Neural Network (FNN) discriminator presented by Lienhard et al.~\cite{Lienhard2022} that optimize the readout pipeline. 

\begin{center}
    \textbf{Access to ADC data traces on IBM Quantum Cloud is restricted. In this paper, we use the dataset of a custom five-qubit chip provided by the authors of~\cite{Lienhard2022}.}
    \vspace{-0.1in}
\end{center}

\begin{figure}[b]
    \vspace{-0.25in}
    \centering
    \includegraphics[width=\linewidth]{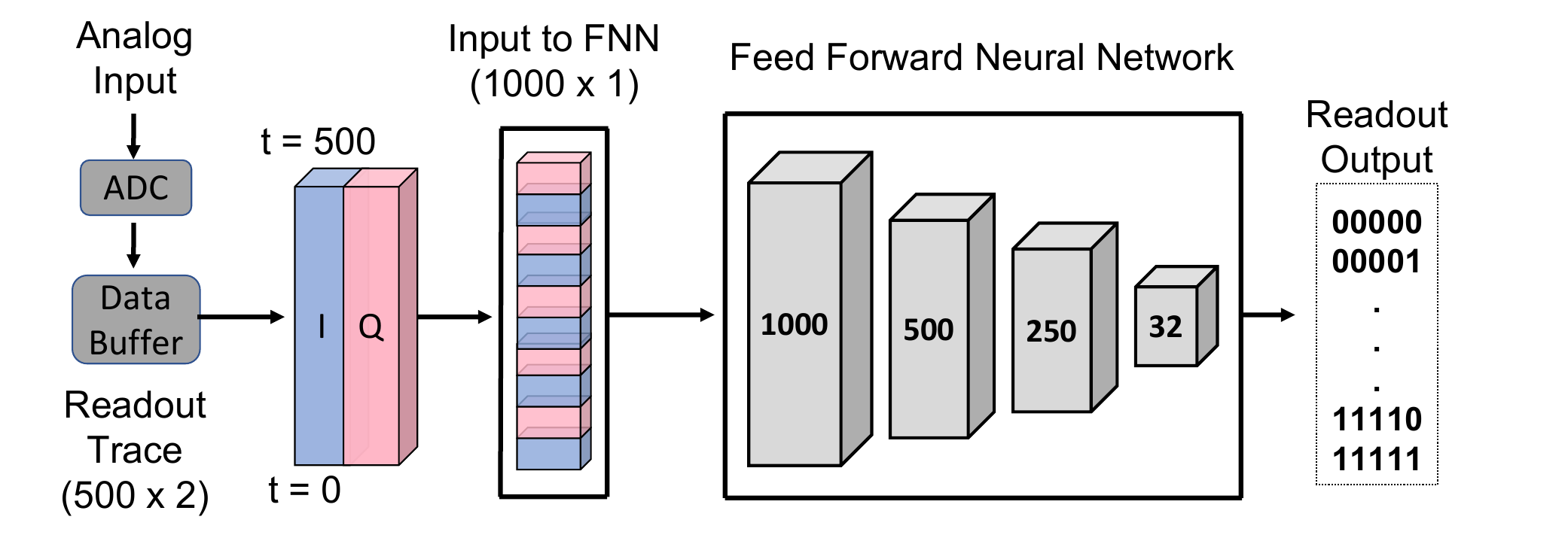}
    \vspace{-0.25in}
    \caption{Overview of the baseline~\cite{Lienhard2022} design.}
    \label{fig:baseline}
\end{figure}

\subsection{Baseline Design}
An overview of the baseline FNN \rev{(1000-500-250-32)} design proposed by Lienhard et al.~\cite{Lienhard2022} is illustrated in Figure~\ref{fig:baseline}. The intermediate-frequency readout signal is digitized and buffered before being sent to software classifier. Every readout trace has 500 elements for both I and Q channels, with each element corresponding to a 2ns interval \rev{(total readout duration is 1$\mu$s)}. The 1000 elements corresponding to a single trace are then used as inputs to an FNN. \rev{The baseline design does not demodulate the readout signal. It uses all samples from the ADC to minimize undersampling, and thus the input layer consists of 1000 neurons.} The FNN has 32 outputs corresponding to a five-qubit system's $2^5$ basis states.



\subsection{Factors affecting single-shot accuracy}
\subsubsection{Relaxation errors}
Figure~\ref{fig:problem}(a) shows examples of how measurements of the excited state undergo relaxations ($1 \rightarrow 0$ transitions) during readout. A significant number of measurements corresponding to `1' end up in the region occupied by the `0' measurements. 
This presents an essential trade-off for the readout protocol: Shorter readout periods will result in fewer relaxation errors but are more challenging to discriminate. Shorter, and thus often stronger measurements can also result in leakage errors~\cite{Sunada2022}. Figure~\ref{fig:problem}(b) shows the quantitative effect of the trade-off between readout duration and accuracy: Classification of the ground state is more accurate than the excited state.

\subsubsection{Readout crosstalk}
Frequency-multiplexed readout (multiple qubits are measured at different frequencies using the same physical channel) is susceptible to a non-ideality such as readout crosstalk, where the state of neighboring qubits affects the readout result. The baseline~\cite{Lienhard2022} showed significant reductions in these errors by using an extensive neural network.

\subsection{Readout latency and hardware complexity}
State-of-the-art cloud-based quantum computers by IBM offer three native discriminators -- a hardware centroid-based discriminator, linear, and quadratic discriminators implemented in software~\cite{ibm_discriminators}. \rev{Software discriminators enable higher accuracy at the cost of increased readout latency and hardware complexity as the data used for discrimination must be transferred to software. We observed that using software classifiers in IBM Manila slows the execution time by 15\% on average. For the baseline, this slowdown will be greater, as raw ADC data must be transferred to software, unlike IBM systems where raw data is integrated into a single value.} For near-term systems, this additional latency affects real-time feedback applications, such as active reset~\cite{Crcoles2021}, system-level metrics such as Circuit-Level Operations per Second (CLOPS)~\cite{clops}. For future applications, such as error correction, software discrimination further increases the overhead of decoding syndromes~\cite{Gokul2022}. The FNN discriminator presented in~\cite{Lienhard2022} -- enabling an improvement of readout fidelity -- has the drawback that hardware implementations of such discriminators are expensive, as shown by Figure~\ref{fig:problem}(c) where even a fraction\footnote{Vivado HLS could not synthesize hardware beyond this size.} \rev{(400-200-100-32)} of the FNN used in~\cite{Lienhard2022} uses 4x more LUTs than available in \rev{a Zynq MPSoC (xczu7ev), a device similar to RFSoC based quantum control platforms like QICK~\cite{stefanazzi2022qick}. RFSoCs integrate ADCs, DACs, an FPGA, and a CPU on the same chip, making them scalable for quantum control applications. Synthesis was performed by Vivado High-Level Synthesis (HLS) with optimized resource usage and a reuse factor of 25}. Thus, even though the FNN discriminator achieves significantly higher multi-qubit single-shot fidelities, \rev{scaling a system that uses such an expensive hardware discriminator is impractical. Even if the FNN discriminator is implemented on hardware with additional engineering efforts and larger FPGAs, having a single FPGA for the readout of just 5 qubits will dramatically increase the cost and complexity of the control hardware as the number of qubits are scaled up.}

Considering the challenges in improving single-shot readout accuracy and the requirement of fast, scalable hardware discriminators, the question we seek to answer is the following:
\begin{tcolorbox}[colback=white!80!black,colframe=white!75!black]
  Can readout be performed with fast and scalable hardware discriminators to minimize readout latency while improving single-shot readout accuracy?
\end{tcolorbox}

To answer this question, we use the FNN discriminator proposed by Lienhard et al.~\cite{Lienhard2022} as a baseline and develop an accurate and hardware-efficient discriminator utilizing the same dataset used by the baseline design.

%% file: inputs/4design-1.tex
\section{Enabling Efficient High-Fidelity Readout }
In this paper, we propose {\em High Fidelity Hardware-efficient Qubit Readout using Low-dimensional Traces} ({\sc herqules}), a holistic architecture that focuses on improving the fidelity, reducing the latency, and enhancing the scalability of superconducting qubit readout.







\subsection{HERQULES Design Overview}



\begin{figure}[t]
    \centering
    \includegraphics[width=0.6\linewidth]{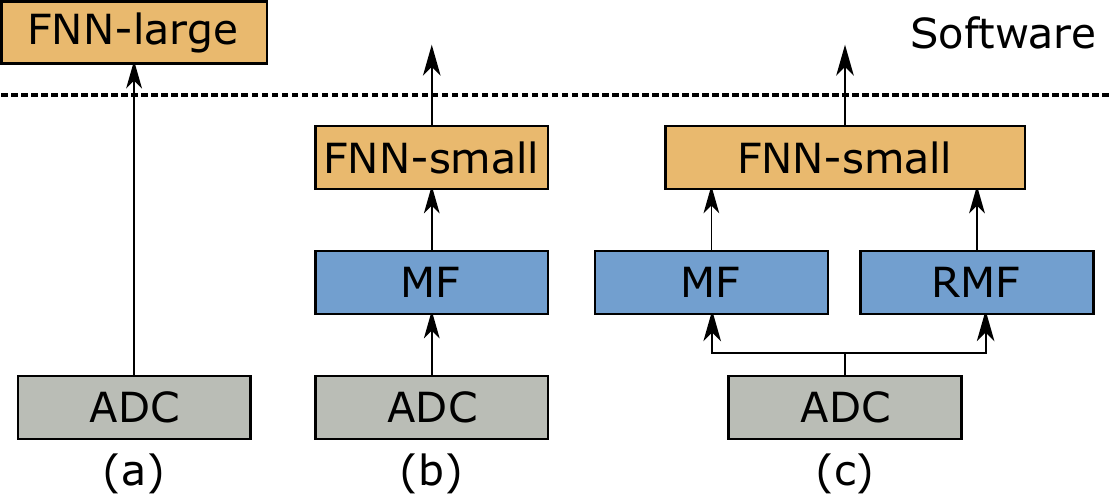}
    \vspace{-0.15in}
    \caption{Overview of pipeline designs; (a) Baseline -- Feed forward Neural network (FNN) on ADC data, without dimensionality reduction; (b) {\sc mf-nn}: dimensionality reduction with matched filter (MF), then as input to FNN (smaller network than the baseline); (c) {\sc mf-rmf-nn}: Additional feature with relaxation matched filter (RMF) over and above {\sc mf-nn}.}
    \label{fig:design_overview}
    \vspace{-0.23in}
\end{figure}

The baseline design~\cite{Lienhard2022} uses a large Feed Forward Neural Network (FNN) trained on the readout traces at intermediate frequencies from the ADC to discriminate qubit state. This architecture learns features corresponding to crosstalk, qubit decay, and other non-idealities to achieve high-fidelity qubit readout. Unfortunately, large FNNs require significant computational and memory resources, forcing designers to run them in software, as shown in Figure~\ref{fig:design_overview}(a). 


To enable efficient design, we propose \ours~that reduces the dimensionality of the readout trace using matched filters (MF) and feeds this data to a significantly smaller, hardware-efficient FNN to infer the qubit states as shown in Figure~\ref{fig:design_overview}(b). A MF reduces the dimensionality of the demodulated time traces to a single dimension with two classes maximally separated. The reduction in the dimension of the input data allows for a smaller FNN, as shown in Figure~\ref{fig:design_overview}(b). Furthermore, \ours~introduces additional classifiers into the readout pipeline to detect qubit relaxation. An additional MF-based discriminator is tasked to catch qubit relaxation (relaxation matched filter (RMF)) as shown in Figure~\ref{fig:design_overview}(c). \ours~shows that by feeding the FNN with other specialized classifiers, such as MFs, we can compensate for non-idealities occurring in frequency-multiplexed qubit readout architectures. 

In the rest of this section, we further describe our design to explain the training (done in software) and the inference (in hardware) processes for \ours.

\begin{figure*}
    \centering
    \includegraphics[width=0.9\linewidth]{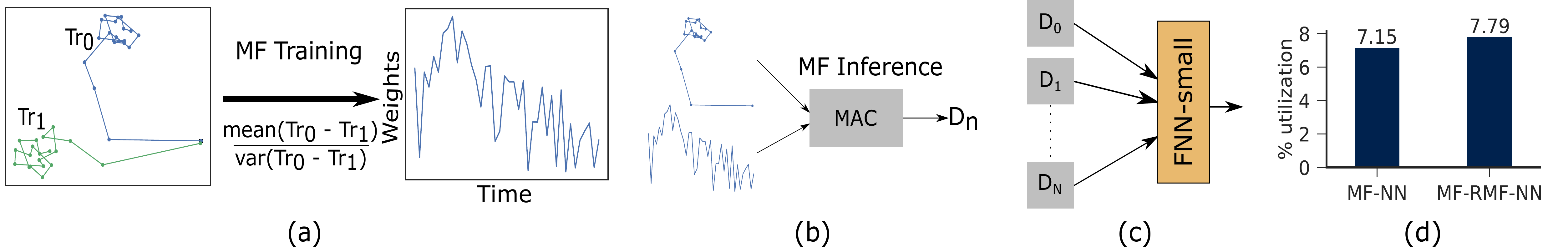}
    \vspace{-0.1in}
    \caption{ (a) Matched Filter (MF) training: Ground state traces ($Tr_{0}$) and excited state traces($Tr_{1}$) are processed to get MF envelopes; (b) The dot product of the traces and MF envelopes is computed using multiply-and-accumulate (MAC) units; (c) The computed dot product is input for a Feed Forward Neural Network (FNN); (d) Comparison of LUT utilization between {\sc mf-nn} and {\sc mf-rmf-nn} (remaining component utilization is relatively low).}
    \label{fig:design}
    \vspace{-0.2in}
\end{figure*}

\subsection{Dimensionality Reduction using Matched Filter (MF)}
The FNN discriminator used every time-bin element of the readout trace received from the qubits as an input feature, resulting in an input layer of 1000 neurons for a 1$\mu$s readout trace~\cite{Lienhard2022}. However, using such a neural network, non-idealities, such as the effects of crosstalk, can be mitigated to improve the combined readout accuracy for a frequency-multiplexed readout scheme. To retain the advantages offered by neural networks in mitigating readout crosstalk while reducing the amount of data used to discriminate the state of a qubit, we use a combination of a kernel for dimensionality reduction and a significantly smaller neural network for mitigating non-idealities. We call this design {\sc mf-nn}.

MFs maximize the signal-to-noise ratio (SNR) for signals with linearly added Gaussian noise and in the absence of state transitions \rev{and have been shown to be optimal for single qubit readout~~\cite{ryan2015tomography} (see Appendix~A)}. Many experimental and commercial superconducting quantum computing systems utilize MFs for qubit-state discriminators~\cite{Bronn2017, presto, Crcoles2021}. As described below, MFs are straightforward in training and inexpensive to implement in hardware, making them ideal for dimensionality reduction with preserving features that maximize the separation between two classes. Note that a Matched Filter (also referred to as Linear Discriminant Analysis~\cite{turin1960introduction, bishop2006pattern}) is a lightweight supervised learning method that requires labeled data. 


\subheading{MF Training}
\rev{The readout signal is 2D (I and Q values) and varies with time.} The MF envelope is derived by computing the ratio of the mean and variance of the difference between the two classes, visualized in Figure~\ref{fig:design}(a). \rev{A separate MF envelope is trained for both I and Q components of the readout signal but for brevity, we refer to both envelopes in the singular.} Since this envelope is unique for every qubit, the number of envelopes that need to be trained for readout scales linearly with the number of qubits. MF's training overhead is significantly lower than the baseline FNN, making \ours~agile and efficient.   

\subheading{MF Inference}
The MF output is essentially a dot product between the trained envelope \rev{$env$} and the \rev{demodulated} trace \rev{$Tr$} being discriminated. \rev{For both I and Q components, the result of the matched filter is thus $\sum_{j \in I,Q}\sum_{t=0}^{T}env_j(t).Tr(t)$, where $T$ is the readout duration. Since the envelopes of both I and Q components are summed together, the final result is a single scalar value.} The MF operation can be implemented in hardware using inexpensive multiply-and-accumulate (MAC) units, as shown in Figure~\ref{fig:design}(b). 

\subsubsection{Feed Forward Neural Network Architecture}
Matched filtering allows a readout time trace to be reduced to a single value. Typically, this value is utilized to discriminate between two states through thresholding. Here, however, we use the filtered values as inputs to a Feed Forward Neural Network (FNN). In general, for the readout of a frequency-multiplexed group of $N$ qubits, $N$ MF outputs form the inputs to a neural network with $N$ neurons in the input layer, as shown in Figure~\ref{fig:design}(c). Since the size of a group of multiplexed qubits is relatively small (5 to 10 qubits), this drastically reduces the size of the neural network and simplifies the design.

\subheading{FNN Training}
The FNN can be trained only {\em after} the MFs for all the qubits in the -frequency-multiplexed group of qubits have been trained. The training of the FNN can then be performed utilizing the same training set as the one used to train the MF. The FNN is trained with a cross-entropy loss with two hidden layers. The overall FNN architecture used is $N\rightarrow2N\rightarrow4N\rightarrow2^N$, where $N$ denotes group size in the frequency-multiplexed group of qubits. 

\subheading{FNN Inference}
The trained FNN can be used to obtain discriminated values for all $N$ qubits in the group. The benefits of using a smaller FNN in hardware are shown in Figure~\ref{fig:design}(d), where the LUT usage on an FPGA for the {\sc mf-nn} design drops to below 8\% for a single FNN catering to five qubits.




\subsection{Improving Readout Fidelity with Qubit Relaxation Detectors}
So far, we have used an MF to reduce the dimensionality of the readout time traces acquired after the ADC and a neural network to mitigate non-idealities. However, Table~\ref{tab:fidelity} shows that the use of a smaller FNN can not match the accuracy of the baseline FNN design~\cite{Lienhard2022}. We observe that the qubit-readout accuracy achieved by the smaller neural network is greatly affected by relaxations ($1\rightarrow0$ transitions). Typically, during measurements, qubits can undergo undesired state transitions. Excited-state relaxation is a stochastic process and its probability increases exponentially in the measurement operation latency. To enhance readout fidelity, we focus on detecting qubit relaxation to reduce the diminishing impact on the qubit-readout accuracy.

\begin{table}[t]
\begin{center}
\begin{small}
\caption{\rev{Qubit-readout accuracy of baseline~\cite{Lienhard2022}, different designs, and cumulative accuracy $F_{5Q}=\sqrt[5]{F_1F_2F_3F_4F_5}$.}}
\label{tab:fidelity}
\setlength{\tabcolsep}{0.05cm} 
\renewcommand{\arraystretch}{1.2}
\vspace{-0.1in}
\scalebox{0.9}{
\begin{tabular}{clccccccc}
\hline
\multicolumn{2}{c}{\textbf{Design}}                              & \textbf{Qubit 1} & \textbf{Qubit 2} & \textbf{Qubit 3} & \textbf{Qubit 4} & \textbf{Qubit 5} & $\mathbf{F_{5Q}}$ & \rev{$\mathbf{F_{4Q}}$} \\ \hline \hline
\multicolumn{2}{c}{\textbf{Baseline}}                            & 0.969            & 0.753            & 0.943             & 0.946            & 0.97  & 0.912  & 0.957            \\ 
\multicolumn{2}{c}{\textbf{{\sc mf}}}                            & 0.968           & 0.734           & 0.891           & 0.934           & 0.956   & 0.892     & 0.937      \\ 
\multicolumn{2}{c}{\rev{\textbf{{\sc mf-svm}}}}                  & 0.968           & 0.738           & 0.895           & 0.928           & 0.953   & 0.892     & 0.936      \\ 
\multicolumn{2}{c}{\textbf{{\sc mf-nn}}}                         & 0.969           & 0.740           & 0.901           & 0.936           & 0.957    & 0.896    & 0.940     \\ 
\multicolumn{2}{c}{\rev{\textbf{{\sc mf-rmf-svm}}}}              & 0.981           & 0.752           & 0.959           & 0.957           & 0.986   & 0.923     & 0.970      \\ 
\multicolumn{2}{c}{\cellcolor{green!25}\textbf{{\sc mf-rmf-nn}}} & \cellcolor{green!25}0.985           & \cellcolor{green!25}0.754           & \cellcolor{green!25}0.966           & \cellcolor{green!25}0.962           & \cellcolor{green!25}0.989    & \cellcolor{green!25}0.927   & \cellcolor{green!25}0.975    \\ \hline
\end{tabular}
}
\end{small}
\vspace{-0.25in}
\end{center}
\end{table}

\subheading{Insight}
Relaxations during readout cause a bias, i.e., the readout fidelity of qubits in the ground (`0') state exceed the one of qubits in the excited (`1') state. This bias has led to work such as invert and measure, where this state-dependent bias is used to improve the overall readout accuracy~\cite{tannu2019mitigating,smith2021qubit,Alexandrou2021,Funcke2022}. However, all prior works focus on reducing the readout bias using statistical techniques applied to a large number of qubit measurements and do not improve the single-shot qubit-readout fidelity, which is essential for quantum error correction. The critical insight of our design is -

\begin{center}
{\em The readout trace of a qubit relaxing during readout differs from error-free traces for the ground and excited states.} 
\end{center}

Using this insight, we train an MF to detect relaxations ($1\rightarrow0$ transitions) during readout to accurately estimate the state of the qubit prior to the measurement.  

\subsubsection{Identifying relaxation traces during training}
Training a machine learning model to detect qubit relaxation in a supervised manner is challenging, as qubit relaxation is an uncontrolled stochastic process. Unlike typical supervised training used for readout classifiers, where a qubit is prepared in a known state to generate a training dataset, creating labeled relaxation traces is implausible. Moreover, the relaxation traces are device dependent. To solve this problem, we propose a computationally efficient and accurate unsupervised algorithm that automatically identifies the relaxations and classifies qubit-readout traces as true ground states (`0'), excited states (`1'), or qubit-state relaxations ($1\rightarrow0$) traces.


{
\setlength{\textfloatsep}{0pt}
\SetAlgoNoLine
{
\begin{algorithm}
\begin{small}
\caption{Identify relaxation traces in training set}
\label{alg:relaxations}
\SetKwInput{KwInput}{Input}                
\SetKwInput{KwOutput}{Output}              
\DontPrintSemicolon
    \KwInput{ $Tr_{in}$: Readout traces from training set }\vspace{0.0 in}
    \KwOutput{ $Tr_{relax}$: Relaxation traces}\vspace{0.05 in}
  \SetKwFunction{relaxations}{getRelaxationTraces}
  \SetKwProg{Fn}{Function}{:}{}
  \Fn{\relaxations{$Tr_{in}$}}
  {
    $Tr_{in}$: $N_{train}$ traces, each trace a 2D vector\;
    $traces_0 = Tr_{in}[0]$ \CommentSty{// (`0') traces}\;
    $traces_1 = Tr_{in}[1]$ \CommentSty{// (`1') traces}\;
    \For{$i$ \textbf{in} $N_{train}$} {
        $mean_0[i] = \mathbf{mean}(traces_0[i])$\;
        $mean_1[i] = \mathbf{mean}(traces_1[i])$\;
    }
    $centroid_0 = \mathbf{mean}(mean_0)$\;
    $centroid_1 = \mathbf{mean}(mean_1)$\;
    $radius = \mathbf{distance}(centroid_0, centroid_1) / 2$\;
    \For{$tr_1$ \textbf{in} $traces_1$} {
        \lIf{$\mathbf{distance}(tr_1, centroid_0) \leq radius$} {\;
        \CommentSty{// Append index of original trace}\;
            \hspace{0.1in}$Tr_{relax}$.append($\mathbf{index}(tr_1)$)
        }
    }
  }
  \KwRet{$Tr_{relax}$}
\end{small}
\end{algorithm}
}
}

We propose Algorithm~\ref{alg:relaxations} to identify relaxation traces. For every qubit-readout trace in the baseline dataset, we have labels '0' and '1.' This labeled dataset is typically generated during the device calibration. The proposed Algorithm~\ref{alg:relaxations}, further refines these labels into trusted ground (`0'), excited (`1'), and qubit-state relaxation ($1\rightarrow0$) time traces without running any additional calibration experiments. For example, we use the baseline dataset consisting of $N_{train}$ traces, with each trace being a 2D matrix (one dimension each for the I and Q components) with a length equal to the readout duration. \rev{From the baseline dataset, Algorithm~\ref{alg:relaxations} estimated 4.3\%, 8.9\%, 11.6\%, 6.5\% of the total traces were relaxation traces for qubits 1, 3, 4, and 5 respectively (the lack of distinguishability results in noisy results for qubit 2).}

\begin{figure}[t]
    \centering
    \includegraphics[width=\linewidth]{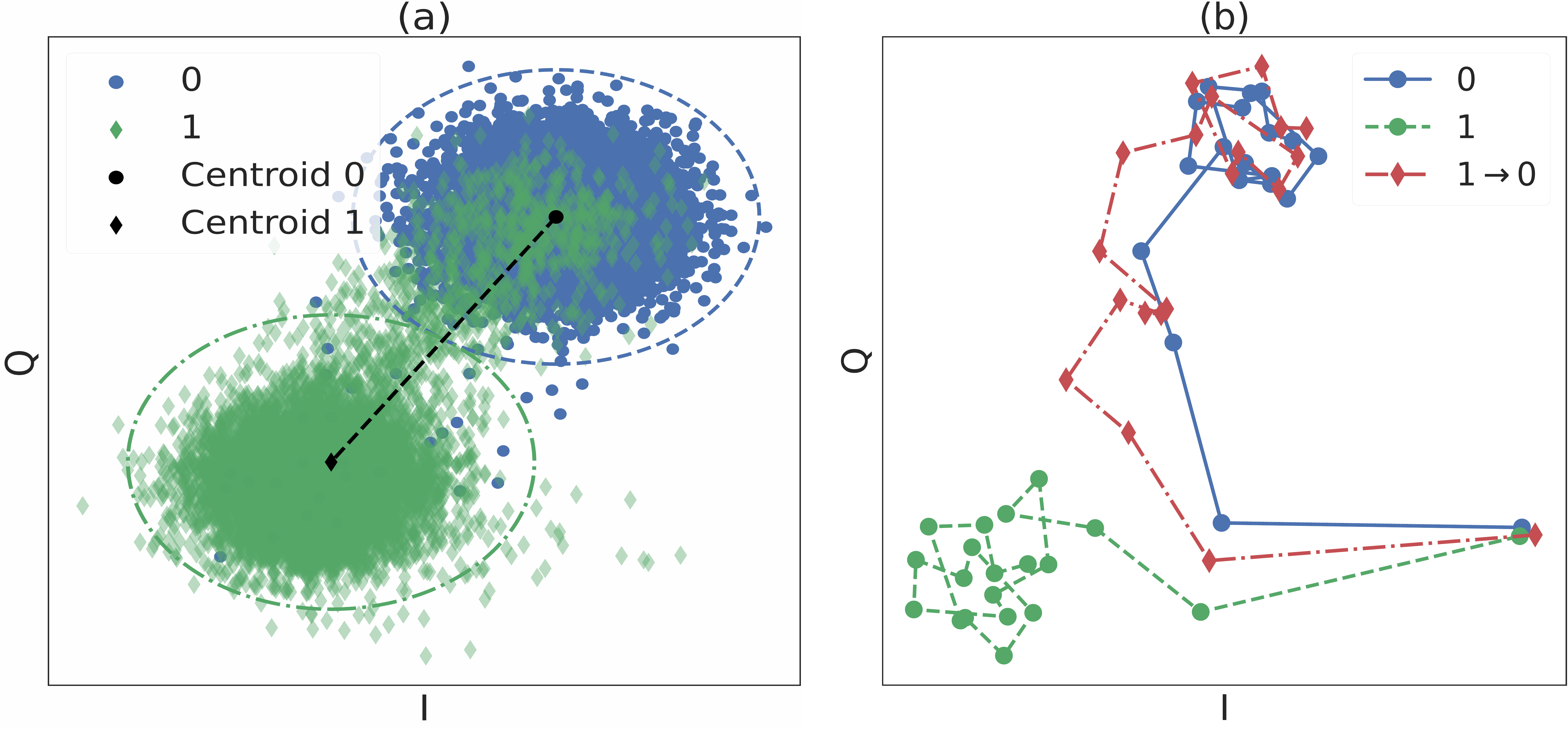}
    \vspace{-0.23in}
    \caption{(a) Plot of I vs. Q values along with centroids of ground and excited-state traces shown as mean-trace values (MTV); (b) The traces corresponding to a ground state (0), excited state (1), and relaxation (1$\rightarrow$0).  
    }
    \label{fig:detect_relax}
    \vspace{-0.22in}
\end{figure}

For labeling relaxation traces, we first reduce the dimensionality of each qubit-readout trace by computing the {\em Mean Trace Value (MTV)}, which is a mean value of I and Q components for the trace.
The key insight behind this algorithm is that traces that relax to the ground state have a \textit{MTV} similar to traces that are correctly identified as `0,' as depicted in Figure~\ref{fig:detect_relax}(a) showing the \textit{MTV} of all traces in the training set. The measurements that are labeled as `1' but lie in the region dominated by the `0' measurements are thus either (a) traces corresponding to relaxations that occurred {\em during} readout, (b) traces corresponding to relaxations that occurred {\em before} readout, or (c) initialization errors. For training purposes, (b) and (c) present the same behavior. To keep the identification simple, we assume that all cases are due to reason (a), even though this may bias the process. Thus, to detect relaxation errors, we define a circular region for both classes, where the centers are the means of all points corresponding to each class (centroids), and the radius is half the distance between the two centroids.



\subsubsection{Adding features for better accuracy}
Once traces corresponding to qubit-state relaxations have been identified, they can be used to train a classifier that can distinguish between relaxations and true `0' traces. Choosing the correct filter/detector is the next step. As shown in Figure~\ref{fig:detect_relax}(b), the time evolution of a trace corresponding to a qubit-state relaxation is different from a trace corresponding to a reliable ground state measurement. We use this observation to create another MF for every qubit in the system, here referred to as a Relaxation Matched Filter (RMF). The RMF is trained using the relaxation traces ($Tr_{relax}$) identified from the training set and the ground-state traces ($Tr_0$) with the formula $\mathbf{mean}(Tr_{relax} - Tr_0) / \mathbf{var}(Tr_{relax} - Tr_0)$. This is thus different from the general MFs, which are trained using the ground and excited-state traces. 

\begin{figure}[b]
    \centering
    \vspace{-0.2in}
    \includegraphics[width=0.92\linewidth]{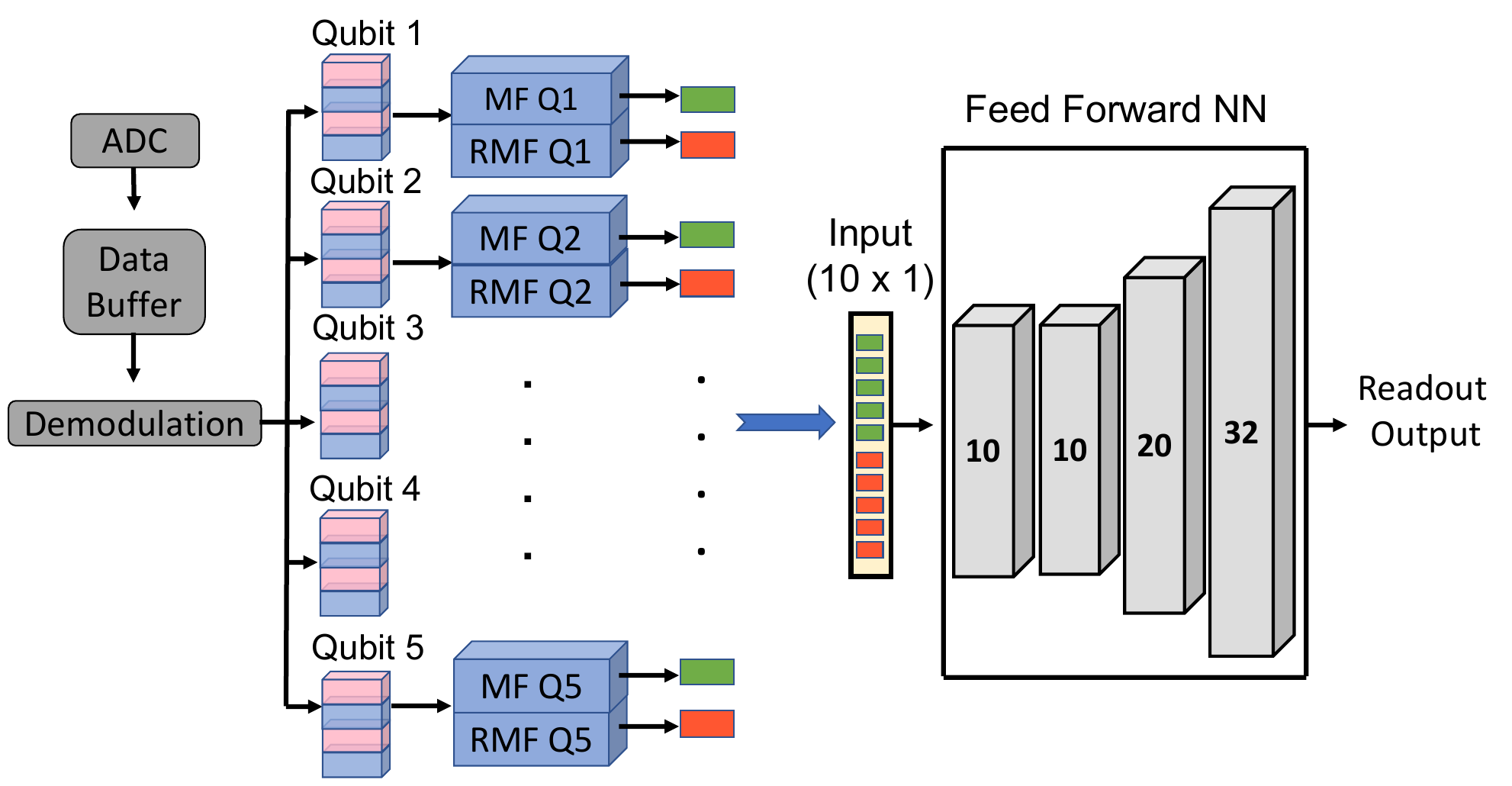}
    \vspace{-0.15in}
    \caption{Proposed {\sc mf-rmf-nn} design overview.}
    \label{fig:MF-RMF-NN}
\end{figure}
The second part of our design involves expanding the inputs of the FNN used in the {\sc mf-nn} design to accommodate the output of the RMF. Since there are $N$ RMFs for $N$ qubits in a frequency-multiplexed group of qubits, the input layer of the FNN increases from $N$ to $2N$ neurons. This intuitively corresponds to adding more features to the input data for the neural network to use, thus changing its role from just identifying crosstalk to using the data from both kinds of MFs to yield an optimal result. Depicted in Figure~\ref{fig:MF-RMF-NN}, we call this design {\sc mf-rmf-nn}. \rev{The {\sc mf-nn} design can also be derived from {\sc mf-rmf-nn} by removing the RMFs and reducing the input layer of the neural network to 5 neurons.}


The benefit in discrimination accuracy by adding the RMF and modifying the role of the FNN is summarized in Table~\ref{tab:fidelity}. For almost all qubits, we see a minimum of 1.5\% increase in qubit-readout accuracy with {\sc mf-rmf-nn}, with qubit 5 reaching 98.9\% accuracy. Qubit 2's lack of improvement can be explained by the fact that there was no separation between the ground and excited states, due to which the algorithm used for determining relaxation traces yielded very noisy results. We achieve substantial gains in accuracy without using a significant amount of hardware -- as shown in Figure~\ref{fig:design}(d), the increase in LUT usage on an FPGA due to an increase in the size of the input layer and additional MFs is marginal, utilization increases from 7.15\% to 7.79\% to support five-qubit readout. Compared to the FNN implemented in the baseline, \rev{which at best requires an entire FPGA}, \ourspace allows high-fidelity and scalable qubit readout. Overall, we see a 1.44\% improvement in cumulative accuracy ($F_{5Q}$) over the baseline, which corresponds to a $\mathbf{16.4\%} (=\frac{92.66-91.22}{100-91.22})$ relative improvement in the readout infidelity. \rev{If qubit 2 is omitted\footnote{The distinguishability of the states of qubit 2 is limited due to the experimental setup~\cite{Lienhard2022}.}, then the improvement to the cumulative accuracy ($F_{4Q}$) is 1.85\% over the baseline, corresponding to a \textbf{42.9\% relative improvement} in inaccuracy. The precision (recall) of the predictions made by {\sc mf-rmf-nn} for the five qubits were 98.6\% (98.4\%), 78.6\% (69.7\%), 97.1\% (96.2\%), 95.5\% (97\%), and 98.8\% (99.1\%) respectively.
}


\begin{figure}[t]
    \centering
    \includegraphics[width=\linewidth]{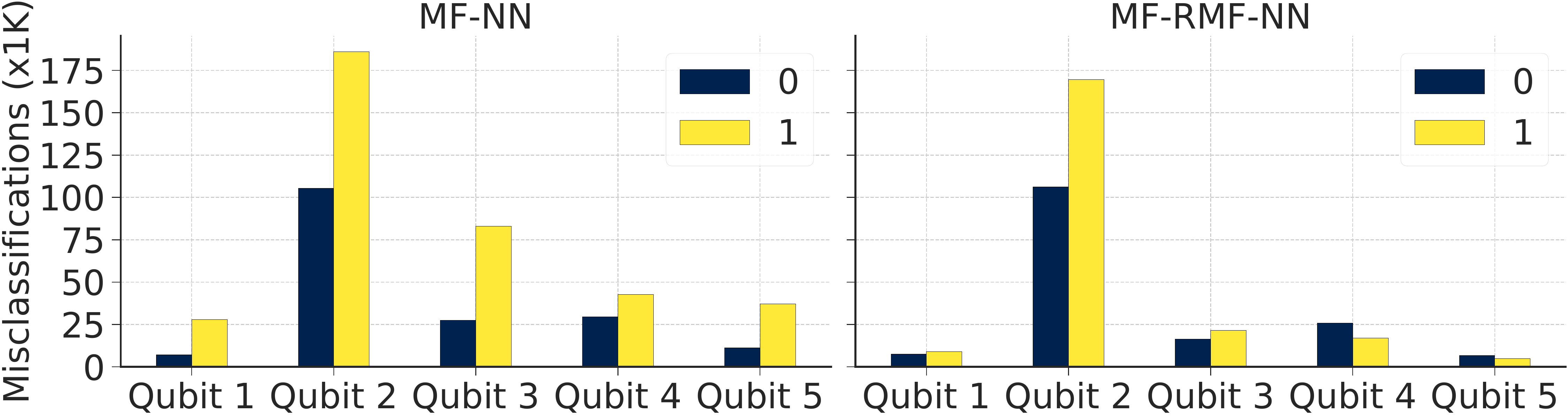}
    \vspace{-0.25in}
    \caption{Comparison of misclassification of ground (`0') and excited (`1') states for all qubits between {\sc mf-nn} and {\sc mf-rmf-nn} designs.}
    \label{fig:rmf_improv}
    \vspace{-0.25in}
\end{figure}
{\color{black}
\subsubsection{Crosstalk mitigation}
To illustrate the benefit of neural networks for qubit-state discrimination, we perform cross-fidelity~\cite{Lienhard2022, Heinsoo2018} studies. Cross-fidelity is a measure of correlations between parallel qubit measurements, thus capturing information on correlated errors occurring due to crosstalk. Cross-fidelity is defined as: $F^{CF}_{ij} = 1 - [P(e_i|0_j) + P(g_i|1_j)], (i \neq j)$; where $0_j$ ($1_j$) is the preparation of the qubit $j$ in the ground (excited) state and $g_i$ ($e_i$) is the measured qubit $i$ in the ground (excited) state. Cross-fidelity captures the crosstalk between readout resonators $i, j$ (since every qubit is connected to its own resonator). Table~\ref{tab:cross-fidelity} shows that using the neural network reduces crosstalk in all cases, especially for a hamming distance of one where readout crosstalk dominates (qubit-qubit interactions dominate errors for hamming distances $2$ and $4$) where the reduction is more than 3x compared to a design that uses a linear Support Vector Machine (SVM) instead of a neural network. Note that supressing crosstalk is essential as correlated errors are hard to detect and correct. 
}
\begin{table}[hpt]
\begin{center}
\rev{
\begin{small}
\caption{Mean absolute values of the cross-fidelity $(F^{CF})$ for different hamming distances (lower is better).}
\label{tab:cross-fidelity}
\setlength{\tabcolsep}{0.05cm} 
\renewcommand{\arraystretch}{1.2}
\vspace{-0.1in}
\scalebox{0.85}{
\begin{tabular}{ccccc}
\hline
\textbf{Design}    & $\langle |F^{CF}_{j=i\pm 1}|\rangle$              & $\langle |F^{CF}_{j=i\pm 2}|\rangle$              & $\langle |F^{CF}_{j=i\pm 3}|\rangle$              & $\langle |F^{CF}_{j=i\pm 4}|\rangle$              \\ \hline \hline
\textbf{Baseline}        & 0.002           & 0.005           & 0.002           & 0.0003          \\
{\sc MF                } & 0.0108          & 0.015           & 0.0021          & 0.0008          \\
{\sc MF-NN             } & 0.0071          & 0.011           & 0.003           & 0.0003          \\
{\sc MF-RMF-SVM        } & 0.011           & 0.0077          & 0.0024          & 0.0006          \\
{\sc \textbf{MF-RMF-NN}} & \textbf{0.0031} & \textbf{0.0062} & \textbf{0.0008} & \textbf{0.0005} \\ \hline
\end{tabular}
}
\end{small}
}
\vspace{-0.22in}
\end{center}
\end{table}

\subsubsection{Why does the RMF improve accuracy?}
By training the RMF to separate qubit-state relaxation traces and ground-state traces, we provide the FNN with additional features of the qubit-readout signal. Figure~\ref{fig:rmf_improv} shows empirically how additional features provided by the RMF help reduce the number of misclassifications made by the {\sc mf-rmf-nn} discriminator for the excited (`1') state for all qubits, which improves the overall readout accuracy. For qubit 2, the information provided by the RMF was noisy and the improvement was marginal.


%% file: inputs/5design-2.tex
\section{Enabling Fast Readout}
Qubit readout is among the slowest operations on superconducting quantum processors. The typical readout duration ranges from 300ns to more than a microsecond~\cite{ibm_systems, weber}. In this section, we discuss how our proposed design can be used to provide more flexibility to the programmer by supporting faster qubit readout after calibration.

\subsection{Reducing readout duration}
Fundamentally, the qubit readout duration depends on factors such as the qubit-resonator coupling -- the stronger the coupling, the faster the readout -- and the resonator linewidth -- a larger linewidth results in faster readout~\cite{Krantz2019}. This has been achieved in practice with the help of Purcell filters~\cite{Bronn2017}. Such circuit-level optimizations are beyond this paper's scope. Here, we focus on signal processing techniques to achieve scalable, fast, and high-fidelity readout.

\subsubsection{Advantages of faster readout}
A key advantage of shorter readout durations is the decreased probability of qubit-state relaxations occurring during qubit readout. Thus, readout circuits are typically designed to ensure that the readout duration is significantly shorter than the qubit coherence time~\cite{Krantz2019}. Other advantages are at the system level -- faster readout improves metrics such as Circuit-level Operations per Second~\cite{clops} and reduces the feedback/Feed Forward latency, which is critical for applications utilizing mid-circuit measurements~\cite{Crcoles2021, hua2022exploiting}. Such applications have resulted in studies to optimize the readout electronics to minimize the Feedback/Feed-Forward latency~\cite{Guo2022, salathe2018low}.

\subsubsection{Signal processing methods}
Qubit-state relaxation errors during readout have resulted in using signal processing techniques such as boxcar filtering to improve and simplify qubit readout~\cite{gambetta2007protocols}. The boxcar filter length is optimized for every qubit. Boxcar filters can also be used in conjunction with MFs to achieve better overall readout accuracies~\cite{Lienhard2022} to shorten the MFs to minimize the state-relaxation probabilities. 

\subsection{Application specific support for fast readout}
The baseline FNN design~\cite{Lienhard2022} cannot support shorter readout traces without an additional training phase because the input layer depends on the initial readout duration. Thus, any change in readout duration alters the architecture of the FNN, requiring the FNN to be trained all over again. In comparison, our proposed design can facilitate the discrimination of short readout traces since using MFs for dimensionality reduction makes the neural network agnostic to the actual qubit readout duration. To simplify the overall system and enable readout of nearly arbitrarily long qubit readout traces, we thus propose the following: Train the MFs and neural networks on the complete readout duration. Once training is full, an iterative sweep can be done on the readout duration to find the shortest time that results in a cumulative accuracy (given by the geometric mean of accuracies of individual qubits) that saturates. For our evaluations, we train our design on the complete readout duration and then reduce the length of the readout traces in the test set to determine the accuracy.

\begin{figure}[t]
    \centering
    \includegraphics[width=\linewidth]{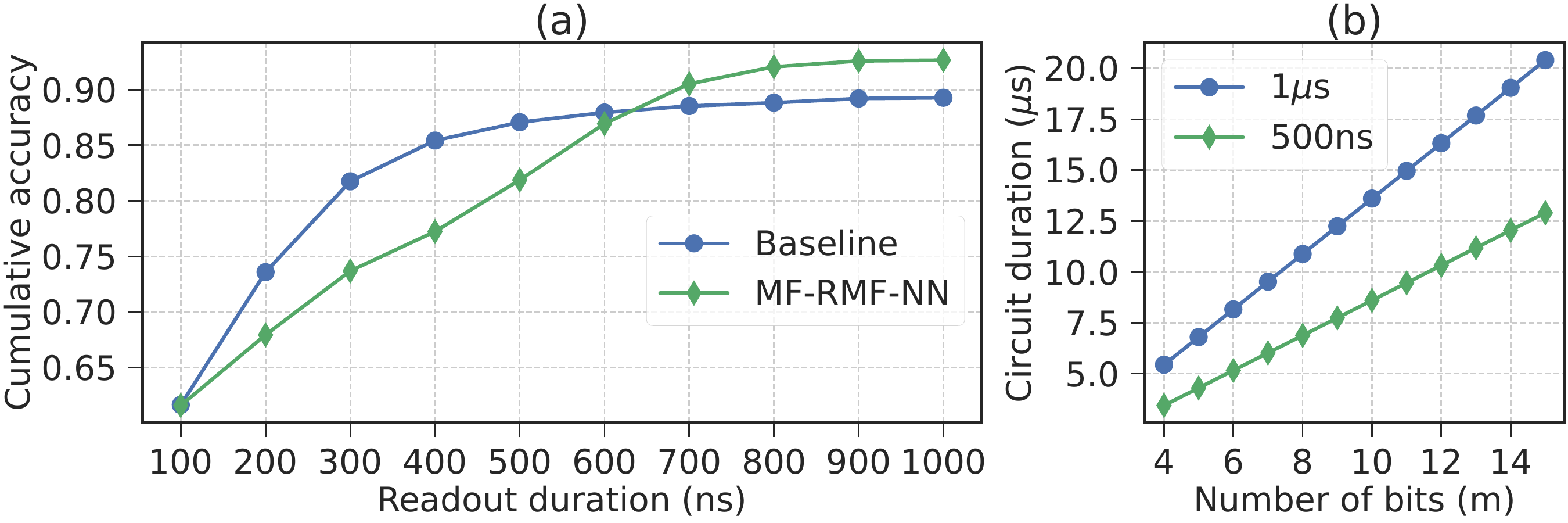}
    \vspace{-0.2in}
    \caption{ (a) Comparison of cumulative readout accuracy vs. readout duration for the baseline and the proposed {\sc mf-rmf-nn} design; (b) Effect of change in readout duration by comparing circuit duration vs. the number of bits for Quantum Phase Estimation (QPE) circuit. }
    \label{fig:fnn_duration}
    \vspace{-0.2in}
\end{figure}


Figure~\ref{fig:fnn_duration}(a) shows how the cumulative readout accuracy of the five-qubit device changes with the readout duration (common for all qubits) for the baseline~\cite{Lienhard2022} and our design {\sc mf-rmf-nn}. 
We see that our design achieves an accuracy exceeding the baseline much sooner \emph{without requiring additional training}.
Table~\ref{tab:fast_readout} shows the accuracy for individual qubits for two different readout durations. This data indicates that some qubits can have shorter readout durations than others -- qubit 5 can be read out twice as fast without a significant drop in accuracy. The cumulative accuracy exceeds the reported baseline accuracy with a 25\% shorter duration of 750ns\footnote{We see a slightly lower qubit-state discrimination accuracy for the baseline due to a difference in the seed value chosen for the neural network. However, this does not change the trend we are showing.}.

\begin{table}[hpt]
\begin{center}
\begin{small}
\vspace{-0.15in}
\caption{\ours~Fidelity vs. Readout Duration}
\label{tab:fast_readout}
\setlength{\tabcolsep}{0.05cm} 
\renewcommand{\arraystretch}{1.2}
\vspace{-0.1in}
\scalebox{0.85}{
\begin{tabular}{cccccccc}
\hline
\textbf{Design}                                                                           & \textbf{Duration}      & \textbf{Qubit 1} & \textbf{Qubit 2} & \textbf{Qubit 3} & \textbf{Qubit 4} & \textbf{Qubit 5} & \textbf{$\mathbf{F_{5Q}}$} \\ \hline \hline
\multirow{3}{*}{\textbf{\begin{tabular}[c]{@{}c@{}}{\sc mf-rmf-nn}\\  (accuracy)\end{tabular}}} & \textbf{1$\mu$s} & 0.985           & 0.754           & 0.966           & 0.962           & 0.989         & 0.927  \\  
      & \textbf{750ns}  & 0.951           & 0.742           & 0.955           & 0.958           & 0.987     & 0.914     \\ 
& \textbf{500ns}  & 0.629           & 0.708           & 0.910           & 0.929           & 0.977      &        0.819               \\ \hline 
\end{tabular}
}
\vspace{-0.15in}
\end{small}
\end{center}
\end{table}

Providing information to the compiler about qubits that can be measured faster than others can be useful for mid-circuit measurement applications. The compiler can map the role of an ancilla qubit, which is being measured more frequently, to physical qubits with shorter readout times.

\subsection{Case study: Quantum Phase Estimation}
To quantitatively show the advantage of fast readout on applications that use mid-circuit measurements, we study the Quantum Phase Estimation (QPE) algorithm. Specifically, we investigate the variant used in~\cite{Crcoles2021}, where mid-circuit measurements are used with conditional gates to reduce the resource overhead of executing the circuit. Such mid-circuit measurements have been used to demonstrate error mitigation strategies such as the bit-flip code~\cite{Rist2020}. Figure~\ref{fig:fnn_duration}(b) shows the effect of halving the readout duration (by using qubit 5 from our results for providing feedback) on the total execution time of an iterative QPE circuit for a different number of bits. While gate latencies do not change, the reduced readout latency enables the QPE circuit to scale better for larger problem sizes.

%% file: inputs/6Methodology.tex
\section{Methodology}

\subheading{Quantum hardware and dataset}
We obtained datasets containing the readout time traces collected directly from the ADC originating from the five-qubit chip used in~\cite{Lienhard2022}. These qubits are read out through a common resonator feedline using frequency-multiplexing. The ADC sampling rate is 500 MSamples/sec, and qubit \rev{relaxation ($T_1$)} times range from 7$\mu$s to 40$\mu$s. The dataset contains readout traces for all $2^5$ basis states of the five qubits, with a total of 50,000 traces per basis state ($32 \times 50000 = 1600000$ traces in total). We fixed the readout duration to 1$\mu$s for all qubits.


    

\subheading{Benchmarks}
We evaluated the effect of the improved readout accuracy achieved by \ours~on the \rev{\texttt{qft-n}, \texttt{ghz-n}, \texttt{bv-n}, \texttt{qaoa-n} NISQ benchmarks with \textit{n} qubits.} 


\subheading{Simulation framework}
We performed noisy simulations of NISQ benchmarks with the Qiskit~\cite{Qiskit} Aer simulator with a noise model derived from the 27-qubit IBM Hanoi backend. To evaluate the effect of readout errors on the logical error rates of surface codes, we used the Stim~\cite{gidney2021stim} stabilizer simulator. 



\subheading{FPGA hardware}
To estimate the FPGA resources needed to implement a neural network, we use a combination of the \texttt{hls4ml}~\cite{fahim2021hls4ml} tool and Xilinx Vivado High-Level Synthesis (HLS). \texttt{hls4ml} can take a neural network model written in frameworks such as Keras, and Pytorch and create an equivalent HLS model that can then be synthesized with Vivado HLS. As the target device, we use the Xilinx Zynq MPSoC \texttt{xczu7ev-ffvc1156-2-i}.

\subheading{Software}
All discriminators are implemented in Python. The PyTorch framework is used to build, train, and test the neural network models for our proposed design as well as the baseline. For each of the 50,000 traces per basis state in the dataset, we use 9,750 traces for training, 5,250 for validation, and 35,000 for testing the model. 

%% file: inputs/7Evaluation.tex
\section{Evaluations}
\begin{figure}[t]
    \centering
    \includegraphics[width=\linewidth]{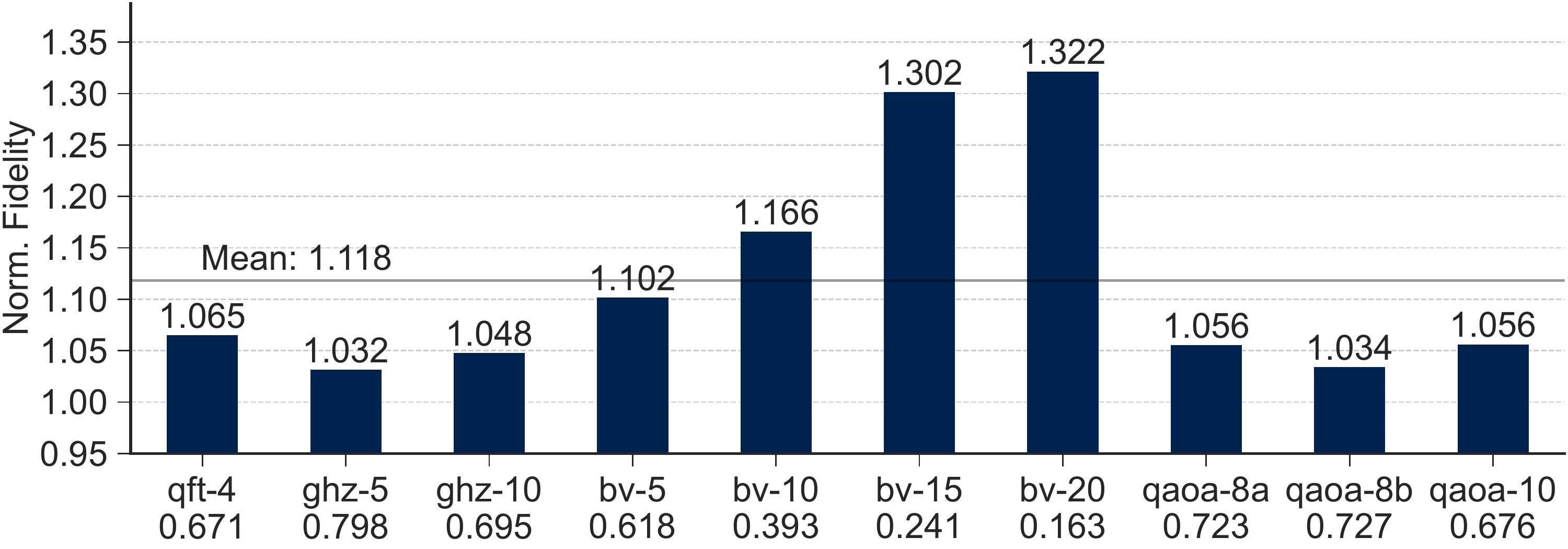}
    \vspace{-0.25in}
    \caption{Normalized fidelity of \ours~with respect to the baseline (baseline fidelity annotated at the bottom) for various NISQ benchmarks.}
    \label{fig:nisq_fidelity}
    \vspace{-0.20in}
\end{figure}
\subsection{NISQ benchmark fidelity}

To evaluate the effect of enhanced qubit-readout accuracies on NISQ benchmarks, we used the Aer simulator from Qiskit~\cite{Qiskit}. We set the readout error for each qubit equal to the geometric mean of individual qubit accuracies for the baseline and our proposed design {\sc mf-rmf-nn}. For the baseline, the cumulative accuracy is \texttt{0.9122}~\cite{Lienhard2022}, and for the proposed {\sc mf-rmf-nn} design, we improve it to \texttt{0.9266}. Other noise parameters, such as single and two-qubit gate errors, are derived from IBM Hanoi. For the QAOA and GHZ benchmarks, we computed the Total Variational Distance (TVD) between the ideal and simulated probability distributions~\cite{hashim2020randomized, lubinski2021application}.

Figure~\ref{fig:nisq_fidelity} shows the normalized fidelity of the {\sc mf-rmf-nn} design with respect to the baseline design. We see a maximum improvement in fidelity for the 20-qubit Bernstein Vazirani benchmark, while other benchmark fidelities improved by at least 3\%. These simulation results demonstrate the benefit of improving readout accuracy in near-term systems. 

\subsection{\rev{Inference latency}}
{
\color{black}
Table~\ref{tab:inf_latency} shows the inference latency of \ours~and the baseline design for a specified Reuse Factor (RF). RF enables the sharing of multipliers to reduce resource utilization (RF of 4 implies that one multiplier is shared for four multiplications). For the baseline, Vivado HLS took an unreasonably long time (exceeding 10 days) for RFs less than 1000. \ours~has orders of magnitude lesser utilization and lower inference latency than the baseline while having a higher accuracy.

}
\begin{table}[hpt]
\begin{center}
\vspace{-0.1in}
\rev{
\begin{small}
\caption{Inference latency and LUT utilization (on Xilinx {\sc xczu7ev}) with different Reuse Factors (RF).}
\label{tab:inf_latency}
\setlength{\tabcolsep}{0.05cm} 
\renewcommand{\arraystretch}{1.2}
\vspace{-0.1in}
\scalebox{0.9}{
\begin{tabular}{cccccc}
\hline
                          & \textbf{\begin{tabular}[c]{@{}c@{}}\ours\\ (RF = 4)\end{tabular}} & \textbf{\begin{tabular}[c]{@{}c@{}}\ours\\ (RF = 64)\end{tabular}} & \textbf{\begin{tabular}[c]{@{}c@{}}Baseline\\ (RF = 200)\end{tabular}} & \textbf{\begin{tabular}[c]{@{}c@{}}Baseline\\ (RF = 500)\end{tabular}} & \textbf{\begin{tabular}[c]{@{}c@{}}Baseline\\ (RF = 1000)\end{tabular}} \\ \hline \hline
\textbf{Latency (cycles)} & 8                                                                     & 21                                                                     & 924                                                                    & 2023                                                                   & 4023                                                                    \\
\textbf{Utilization (\%)} & 7.79                                                                  & 7.24                                                                   & 468.64                                                                 & 266.86                                                                 & 216.72                                                                  \\ \hline
\end{tabular}
}
\vspace{-0.2in}
\end{small}
}
\end{center}
\end{table}


\subsection{Impact on quantum error correction}
\begin{figure}[t]
    \centering
    \includegraphics[width=0.8\linewidth]{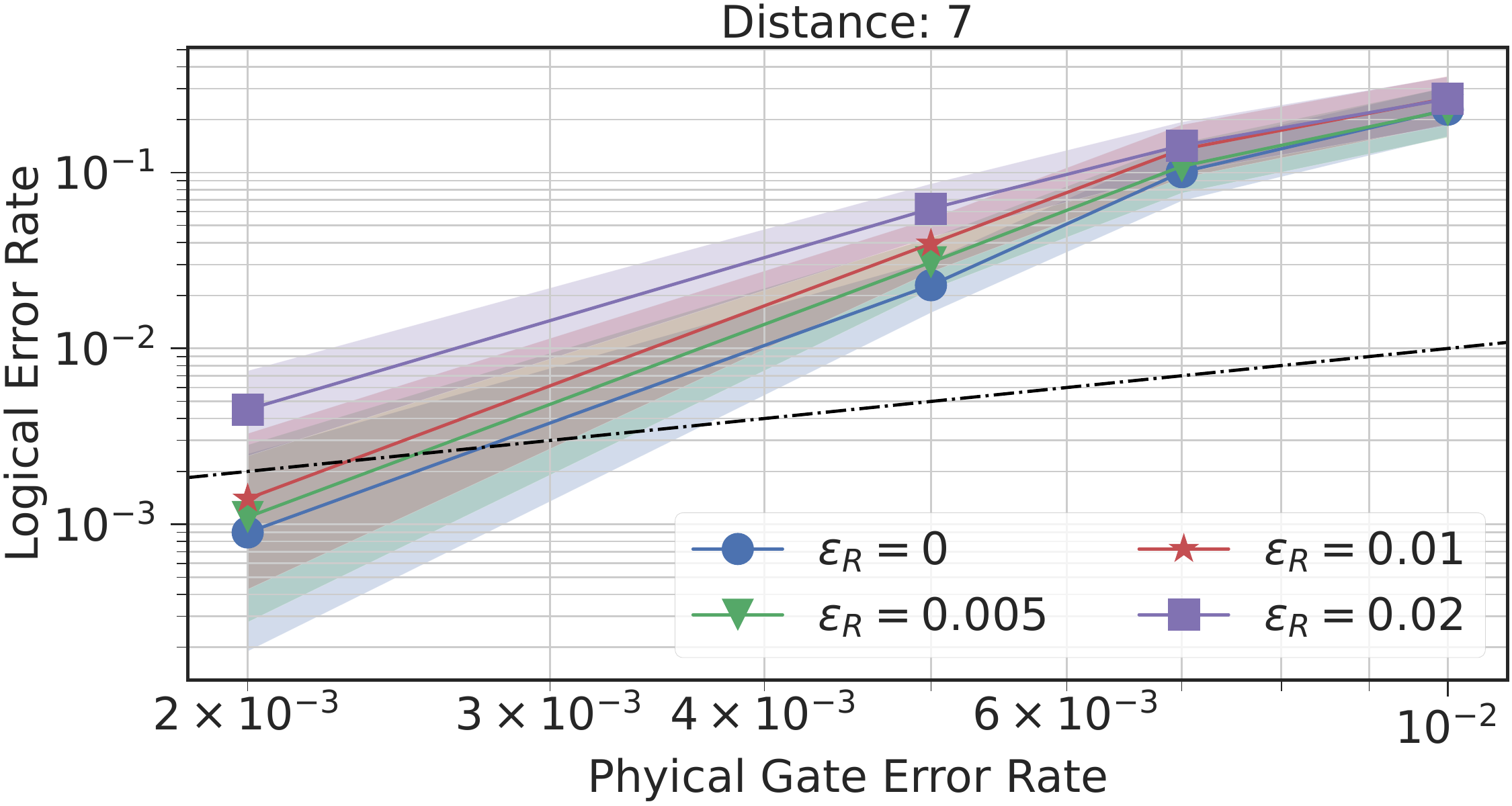}
    \vspace{-0.15in}
    \caption{\rev{Logical error rate per round vs. physical gate error rate for surface code of distance 7 for different values of the averaged readout error ($\epsilon_R$). The logical error rate is equal to the physical error rate at the dash-dot line.}
    }
    \label{fig:ftqc}
    \vspace{-0.15in}
\end{figure}

\subheading{Effect of readout errors}
The unreliability of today's NISQ-era quantum computational systems necessitates using Quantum Error Correction (QEC) codes to increase their usefulness. Surface codes~\cite{fowler2009high} are among the most promising QEC protocols. To evaluate the effect of readout error on surface codes and to show that reducing readout errors is just as important as reducing the physical gate error rate to the efficacy of the surface code, we use the Stim~\cite{gidney2021stim} stabilizer simulator with different values of average readout error rate ($\epsilon_R=$ 0\%, 0.5\%, 1\%, and 2\%). As shown in Figure~\ref{fig:ftqc}, \rev{a 1\% increase in $\epsilon_R$ can cause the logical to surpass the physical gate error rate, which undermines the protection promised by QECs.}


\subheading{Scalability}

\begin{figure}[t]
    \centering
    \includegraphics[width=\linewidth]{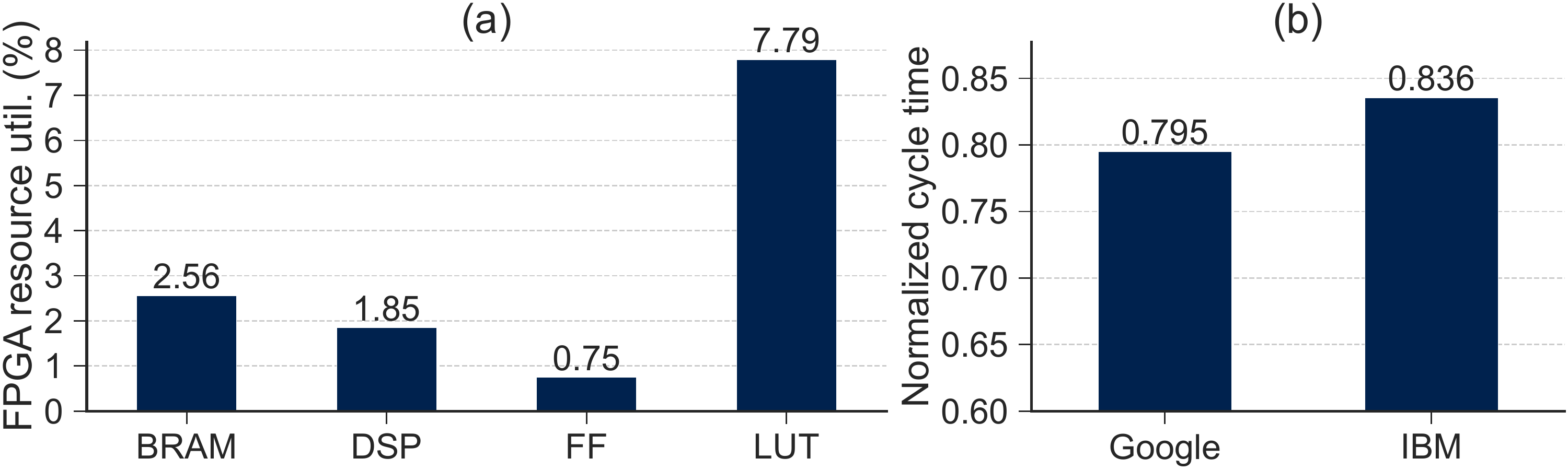}
    \vspace{-0.2in}
    \caption{(a) FPGA resource utilization for \ours; (b) Reduction in the syndrome generation cycle time (lower is better) for a surface-17 circuit~\cite{versluis2017scalable} with gate durations from different systems (Google, IBM).}
    \vspace{-0.2in}
    \label{fig:fpga}
\end{figure}

The FNN used by the baseline design~\cite{Lienhard2022} cannot be efficiently implemented in hardware, making its use for the surface code impractical. This is because the signal output stream from the ADC for every group of multiplexed qubits needs to be transferred to software for discrimination, thus requiring a tremendous amount of bandwidth for large numbers of qubits while increasing the latency of discrimination. In comparison, \ours~can be efficiently implemented in hardware, with resource utilization numbers shown in Figure~\ref{fig:fpga}(a). \rev{Assuming that 80\% of resources are available for readout on an RFSoC controller like QICK~\cite{stefanazzi2022qick}, a single RFSoC can potentially read out more than 50 qubits with extremely low latency. This is in stark contrast to the baseline design which at best can read out a single group of multiplexed qubits per RFSoC. FPGAs such as the Virtex Ultrascale+ have larger fabrics, but do not offer the scalability advantages and cost-effectiveness offered by RFSoCs~\cite{stefanazzi2022qick, presto} for quantum control applications.}

\subheading{Effect of fast readout}
The error-correction cycles of the surface code may benefit from the flexibility to alter the readout duration for each qubit individually. Reducing the readout duration for all qubits while preserving the cumulative accuracy helps shorten the error-correction cycle's duration. This complements other schemes that aim to make error-correction cycles more scalable~\cite{versluis2017scalable}. As shown in Figure~\ref{fig:fpga}(b), reducing the readout duration by 25\% results in a significant reduction in the syndrome cycle time for state-of-the-art quantum processing systems. For processors with faster gates~\cite{weber}, the effect of a shorter readout duration is more pronounced. 

\begin{figure}[b]
    \centering
    \vspace{-0.15in}
    \includegraphics[width=\linewidth]{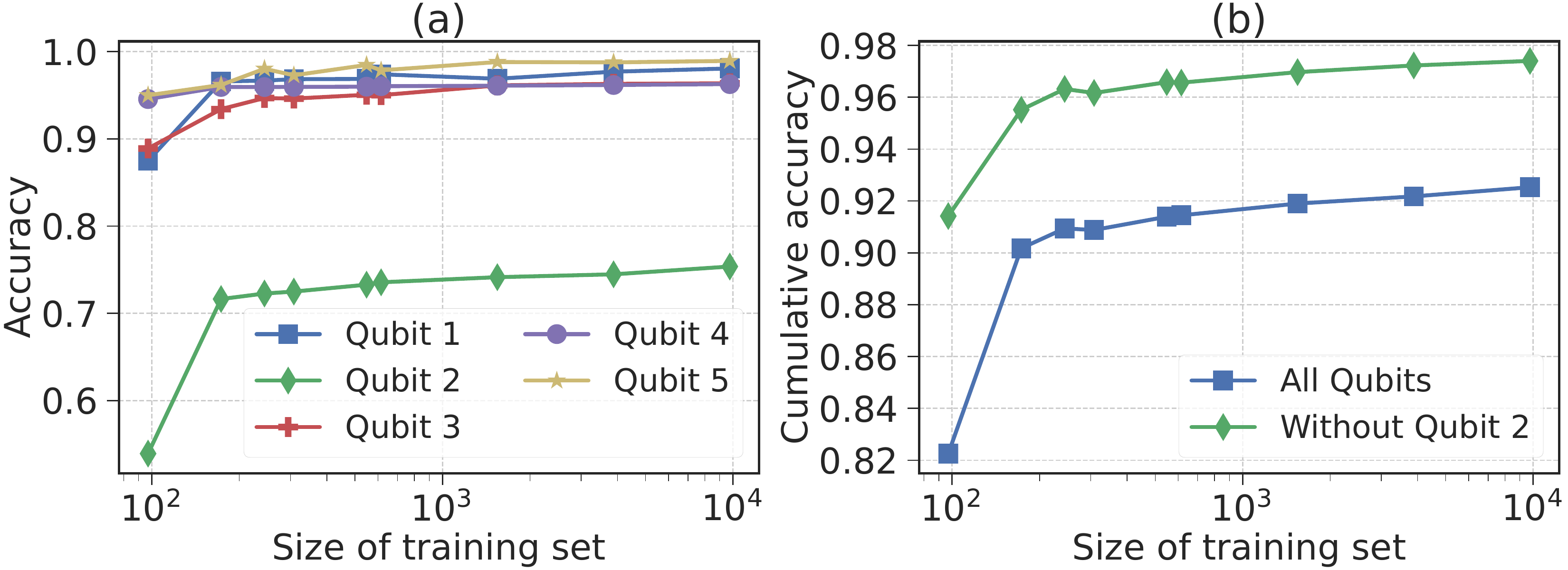}
    \vspace{-0.2in}
    \caption{(a) Accuracy of discrimination of individual qubit in the test set vs. number of training samples used for training;
    (b) Cumulative accuracy of discrimination of all five qubits together in the test set vs. number of training samples used for training.}
    \label{fig:training_curve}
\end{figure}

\subsection{Sensitivity of FNN training in HERQULES}

To evaluate the effect of the training set size on the test accuracy and to ensure that our proposed design is not overfitting the data, we swept through different sizes of the training set (up to a maximum size of 9,750 traces) and determined the accuracy on the test set. For every size of the training set smaller than 9,750, we shuffled the training set. Figure~\ref{fig:training_curve} shows the training curve for (a) individual qubit accuracies and (b) cumulative accuracies given by the geometric mean of the fidelities of $N$ individual qubits ($\textrm{Cumulative accuracy} = \sqrt[N]{F_1F_2...F_N})$. 
With all qubits, the increase in accuracy is 0.77\% when the size of the training set is increased from about 1,500 to 9,750. Without qubit 2, the improvement is 0.4\% for the same interval.

\subsection{Training overhead}

Another advantage of using a much smaller neural network for discrimination is the reduction in total training time.
Table~\ref{tab:training_time} shows the difference in training time between the baseline design and {\sc mf-rmf-nn}. 
Training was done on an AMD EPYC server SoC with 32 physical cores. Shorter training times directly help calibration routine optimizers like SNAKE~\cite{klimov2020snake} since calibration of control parameters for optimal operation of quantum computers is a lengthy process (about four hours for the Sycamore processor~\cite{Arute2019}). 
\begin{table}[hpt]
\begin{center}
\vspace{-0.15in}
\begin{small}
\caption{Total training time for different discriminators.}
\label{tab:training_time}
\setlength{\tabcolsep}{0.05cm} 
\renewcommand{\arraystretch}{1.2}
\vspace{-0.1in}
\scalebox{0.9}{
\begin{tabular}{  c  c  c  c  c }
\hline

    & \textbf{~Baseline~\cite{Lienhard2022}~}  & \textbf{~{\sc mf-rmf-nn}~}  & \textbf{~{\sc mf-nn}~}  & \textbf{~{\sc mf}~} \\ \hline \hline
\textbf{Training time (minutes)}  & ~38  & ~19 & ~17 & ~3  \\ \hline
\end{tabular}
}
\vspace{-0.2in}
\end{small}
\end{center}
\end{table}

%% file: inputs/8Discussion.tex
\section{Discussion}

\rev{To the best of our knowledge, \ours~is the first work that improves qubit readout accuracy \textbf{and} scalability of control hardware needed for readout. Compared to the baseline state-of-the-art qubit state discriminator~\cite{Lienhard2022}, \ours~achieves better accuracy while enabling the readout of potentially an order of magnitude more qubits per FPGA/controller. Additionally, \ours~also enables faster and variable latency readout depending on the application.}

Current FPGA/Radio-Frequency System-on-Chip (RFSoC) control hardware frameworks -- RFSoC are chips that comprise CPUs, ADCs, and DACs along with an FPGA -- such as Qubic~\cite{Xu2021}, Presto~\cite{presto}, and QICK~\cite{stefanazzi2022qick} support most of the standard signal processing blocks in our proposed readout pipeline. For example, demodulation and averaging filters are already supported in QICK, while Presto supports MFs. Thus, the only non-conventional block we add to the readout pipeline is the FNN, which can be easily implemented in hardware due to its small footprint. 



Using FNNs for qubit readout also opens possibilities for leveraging existing research on neural network accelerators for quantum control hardware, especially as the number of qubits is scaled up. The neural network in \ours~can be scaled up in two possible ways -- 1) by having independent FNNs for every group of multiplexed qubits or 2) by having a shared FNN among all qubits.
It is possible for the latter to be more resource efficient while providing better accuracy for all qubits. Still, the output layer (which uses softmax activation) can become prohibitively large for such an architecture as the size of the output layer increases exponentially ($2^{mN}$) with the number of qubits. This could be addressed by partitioning the shared FNN between hardware and software (the on-chip CPU on RFSoCs). The accelerator architectures proposed to improve the utilization of FPGA resources and the efficiency of the neural network in FPGAs~\cite{Shen2017, Ma2017, ding2017circnn} and general frameworks to implement neural network accelerators in FPGAs~\cite{sharma2016high, Wang2016, Fowers2018, Sun2022} can thus be applied for qubit readout.

%% file: inputs/9Conclusion.tex
\section{Related Work}
Recent Readout Error Mitigation (REM) techniques utilize the development in both hardware and software to improve single-shot and multi-shot readout fidelity. 





\subheading{Measurement and Post-processing methods}
Statistical-based approaches mitigate readout errors by approximating the relation between output state probabilities with expected correct outcomes through an error response matrix~\cite{Bravyi2021, ibm_rem}, or a neural network~\cite{Kim2022}. Bit-flip averaging~\cite{Smith} and corresponding probability distribution of measurements \cite{Alexandrou2021} help reduce biased errors. The invert and measure approach~\cite{tannu2019mitigating} uses a relative state-dependent bias to mitigate measurement errors. Jigsaw~\cite{das2021jigsaw} employs measurement subsetting to reduce the impact of readout crosstalk errors for NISQ applications.

\subheading{REM based on Software}
Active detection of readout errors uses ancillary qubits to encode the data qubits, prior to readout, into a bigger multi-qubit array, and mitigate them on a shot-by-shot basis~\cite{Hicks2022}. The appropriate readout pulse length can also significantly improve the qubit state assignment fidelity~\cite{Heinsoo2018}. Employing machine learning techniques through clustering~\cite{Magesan}, classifier~\cite{DisQ} based on output state probabilities, and Feed Forward Neural Network~\cite{Lienhard2022} for the discriminator shown to capture the non-idealities, significantly improved multi and single-qubit assignment fidelities.






\section{Conclusion}
In this paper, we present \ours, a high-fidelity readout discriminator for superconducting qubits with a scalable, hardware-efficient architecture. \ours~achieves up to 2\% improvement (97\%~$\longrightarrow$~99\%) in single-qubit readout accuracy compared to the state-of-the-art baseline design while using a small and lightweight neural network that can be easily implemented on current and future hardware platforms. \ours~can thus be implemented on conventional FPGA control hardware used in commercial and experimental quantum computers to enable low-latency discrimination with very high accuracy. \ours~also enables faster readout without the requirement of an additional training phase -- the cumulative accuracy achieved by \ours~outperformed the baseline design at a 25\% shorter readout duration.
We envision rich architectural design space to enable readout pipelines with higher fidelity and low latency readout and features that can be used to allow for fault-tolerant quantum computers. 